\documentclass[aps,10pt,showpacs,superscriptaddress,noemail,groupedaddress,nofootinbib,preprintnumbers]{revtex4}  
\usepackage{color}
\usepackage{graphicx}  
\usepackage{dcolumn}   
\usepackage{bm}        
\usepackage{amssymb}   
\usepackage{hyperref}
\usepackage{multirow}
\usepackage{ulem}
\usepackage{epsfig}
\usepackage{amsthm}
\usepackage{amsmath}
\newcommand\Eq[1]{Eq.~(\ref{eq:#1})}

\newcommand\Fig[1]{Fig.~\ref{fig:#1}}

\definecolor{pink}{rgb}{1.0, 0.65, 0.79}

\definecolor{evan}{rgb}{0.25, 0.6, 0.25}

\newcommand{\Oh}{\ensuremath{\mathcal{O}_h}}

\newcommand{\berkeley}{
	Department of Physics,
	University of California,
	Berkeley, CA 94720, USA
	}
\newcommand{\jlab}{
	Theory Center, 
	Thomas Jefferson National Accelerator Facility, 
	Newport News, VA 23606, USA
	}
\newcommand{\lbl}{
	Nuclear Science Division,
	Lawrence Berkeley National Laboratory,
	Berkeley, CA 94720, USA
	}
\newcommand{\llnl}{
	Physics Division,
	Lawrence Livermore National Laboratory,
	Livermore, CA 94550, USA
	}
\newcommand{\wm}{
	Department of Physics,
	The College of William \& Mary,
	Williamsburg, VA 23187, USA
	}

\begin{document}

\preprint{LLNL-JRNL-674381, NT@WM-15-13, JLAB-THY-15-2116}

\title{Two-Nucleon Higher Partial-Wave Scattering from Lattice QCD}

\author{Evan~Berkowitz}
\affiliation{\llnl}
\author{Thorsten~Kurth}
\affiliation{\lbl}
\affiliation{\berkeley}
\author{Amy~Nicholson}
\affiliation{\berkeley}
\author{B\'{a}lint~Jo\'{o}}
\affiliation{\jlab}
\author{Enrico~Rinaldi}	
\affiliation{\llnl}
\author{Mark~Strother}
\affiliation{\berkeley}
\author{Pavlos~M.~Vranas}
\affiliation{\llnl}
\author{Andr\'{e} Walker-Loud}
\affiliation{\jlab}\affiliation{\wm}

\collaboration{CalLat, The California Lattice Collaboration}

\date{\today}

\begin{abstract}
We present a determination of nucleon-nucleon scattering phase shifts for $\ell\geq0$. 
The S, P, D and F phase shifts for both the spin-triplet and spin-singlet channels are computed with lattice Quantum ChromoDynamics.
For $\ell > 0$, this is the first lattice QCD calculation using the L\"{u}scher finite-volume formalism.
This required the design and implementation of novel lattice methods involving displaced sources and momentum-space cubic sinks.
To demonstrate the utility of our approach, the calculations were performed in the SU(3)-flavor limit where the light quark masses have been tuned to the physical strange quark mass, corresponding to $m_\pi{=}m_K{\approx}800$~MeV. In this work, we have assumed that only the lowest partial waves contribute to each channel, ignoring the unphysical partial wave mixing that arises within the finite-volume formalism. This assumption is only valid for sufficiently low energies; we present evidence that it holds for our study using two different channels. 
Two spatial volumes of $V \approx (3.5 \rm{fm})^3$ and $V \approx (4.6 \rm{fm})^3$ were used.
The  finite-volume spectrum is extracted from the exponential falloff of the correlation functions. Said spectrum is mapped onto the infinite volume phase shifts using the generalization of the  L\"uscher formalism for two-nucleon systems.  
\end{abstract}


\maketitle


\section{Introduction\label{sec:intro}}
Understanding low-energy nuclear physics directly from the underlying theory of strong interactions, Quantum ChromoDynamics (QCD), remains a primary goal of nuclear physicists.
The motivation can be broadly separated into two categories: obtaining a quantitative description of basic nuclear physics directly from QCD and probing the limits of the Standard Model and its fundamental symmetries through precision low-energy experiments in nuclear environments.
In both cases, there are substantial international experimental efforts planned or underway which require a quantitative understanding of QCD.

The basic interactions of two nucleons (NN) and nuclei are well measured and have led to a variety of precise theoretical descriptions ranging from phenomenological models to effective field theories (EFT).
These over-constrained NN interactions are then used to predict properties of light nuclei using a variety of methods such as multi-nucleon EFT~\cite{Weinberg:1990rz,Bedaque:2002mn,Epelbaum:2008ga},
harmonic oscillator based effective theory~\cite{Haxton:2002kb,Haxton:2006gw},
no-core shell model~\cite{Barrett:2013nh,Navratil:2007we} and 
Green's Function Monte Carlo~\cite{Gezerlis:2013ipa,Carlson:2014vla}.  
For light nuclei, the NN interactions dominate the nuclear structure, but the three-body nuclear force is necessary for accurate comparisons with the measured values~\cite{Pieper:2001mp,Hammer:2012id}.
A recent, exciting development is the use of lattice field theory to regularize the two- and three-nucleon EFT and predict properties of light nuclei, such as the recent computation of the Hoyle State~\cite{Epelbaum:2011md}.

All of these impressive theoretical applications rely upon experimental input of the nuclear interactions.
While this is achievable precisely for the low-energy NN interactions, there are very few constraints on the three and higher nucleon forces.
Determining these interactions directly from QCD is a multifaceted problem.
At the core of this challenge is the non-perturbative nature of QCD which requires a numerical approach at low energy.
Lattice QCD (LQCD) is the discretized version of QCD in a finite Euclidean volume.
It is the only known tool to compute QCD correlation functions in the infrared for which no uncontrolled approximations are necessary.
Using LQCD combined with EFT (see for example Refs.~\cite{Barnea:2013uqa,Kirscher:2015yda}), we will be able to determine the elusive few-body nucleon interactions directly from QCD, relevant for example for the upcoming experiments at FRIB~\cite{Balantekin:2014opa}, designed to study neutron-rich nuclei.
We will also be able to compute the hyperon-nucleon interactions, which are extremely challenging to measure experimentally due to the rapid \textit{weak} decay of the hyperons~\cite{Nemura:2008sp,Beane:2010hg,Inoue:2010es,Beane:2012ey}.
These calculations will be relevant for the experiments planned at the FAIR, JLab and J-PARC facilities.  There are even studies of hyper-nuclei using heavy ion collisions at RHIC~\cite{Adamczyk:2014vca} and the LHC~\cite{Adam:2015nca}.

LQCD is also necessary to compute one, two and few-body nuclear matrix elements such as the scalar matrix elements needed for direct dark matter detection and the electroweak matrix elements which govern nuclear interactions and decays.
Two recent examples of these include the $N\rightarrow N\pi$ parity violating process~\cite{Wasem:2011zz} and the parity conserving $np \to d \gamma$ rate~\cite{Beane:2015yha}.
It is known that, due to technical complications, calculations of matrix elements involving multi-particle states often receive $\mathcal{O}(1)$ corrections from the finite volume~\cite{Lellouch:2000pv,Briceno:2014uqa,Briceno:2015csa}.
In order to control these corrections, the two-particle phase shifts and their derivatives must be determined.
See Ref.~\cite{Briceno:2015dca} for a very nice demonstration of this technology for the case of $\pi\gamma\to\pi\pi$.

Tremendous progress has been made in performing LQCD calculations of two-meson interactions~\cite{Dudek:2012gj,Dudek:2012xn,Pelissier:2012pi,Lang:2014tia,Bali:2015gji,Bulava:2016mks,Guo:2016zos,Feng:2010es,Beane:2011sc}.  These calculations use the L\"{u}scher formalism~\cite{Luscher:1986pf,Luscher:1990ux, Rummukainen:1995vs, Kim:2005gf, Christ:2005gi, He:2005ey, Briceno:2012yi, Hansen:2012tf,  Briceno:2014oea} to relate energy levels in a finite periodic volume to the infinite volume scattering phase shifts.
More recently, this has been extended to include coupled channels such as the $\pi K$--$\eta K$ system~\cite{Dudek:2014qha,Wilson:2014cna} and the $\pi\pi$--$KK$ $I=1$ channel~\cite{Wilson:2015dqa}.
In contrast, the NN system is much more challenging to study for a variety of reasons, see for example~\cite{Walker-Loud:2014iea,Briceno:2014tqa}.
These calculations have been limited to $S$-wave interactions and bound states~\cite{Fukugita:1994ve, Beane:2006mx, Beane:2013br, Beane:2011iw, Aoki:2012tk,Yamazaki:2011nd, Beane:2012vq, Yamazaki:2012hi,Yamazaki:2015asa}.%
\footnote{There has been one exploratory study of higher partial wave NN interactions in Ref.~\cite{Murano:2013xxa}.  This calculation used the \textit{so-called} potential method~\cite{Ishii:2006ec} which suffers from additional systematics that have not yet been demonstrated to be under control. 
To date, there has been one quenched $I=2\ \pi\pi$ comparison~\cite{Kurth:2013tua}.
}
The methodology for determining three-body forces from LQCD calculations is still being developed~\cite{Hansen:2015zga, Meissner:2014dea, Hansen:2014eka, Briceno:2012rv, Polejaeva:2012ut,Doi:2011gq}.
However, it is evident that to reliably extract three and four-body forces a precise determination of two-body scattering parameters is needed, including but not limited to $\ell=0$ partial waves.

In this work, we present a calculation of higher partial wave scattering in the NN system.  In particular, we have computed S,P,D and F partial waves in both isosinglet and isotriplet channels using the NN generalization~\cite{Briceno:2013lba} of L\"uscher's formalism.
Given the complex nature of this problem, this exploratory calculation was performed at the $SU(3)$ flavor symmetric point with $m_\pi \sim800$~MeV, enabling us to explore the implementation of our new method and demonstrate its feasibility with relatively little investment of computing time. We have also simplified the L\"uscher formalism by ignoring mixing from higher partial waves contributing to a given cubic irrep. This mixing is kinematically suppressed at sufficiently low scattering energies, however, for the range of energies we explore the assumption that they do not contribute significantly may require further investigation. Some evidence that the assumption is valid can be obtained by comparing different cubic irreps coupling to the same partial wave, and is presented in Section~\ref{sec:results}. This work is an extension of a previous determination of the NN $S$-wave interactions~\cite{Beane:2013br} on the same LQCD gauge configurations.

\section{Improved Two-Nucleon Interpolating Fields \label{sec:new_ops}}
We have performed the calculations with the isotropic Wilson lattices generated by the JLab/W\&M group, with a lattice spacing of $b=0.145(2)$~fm and two spatial extents of $\text{L}/b=24$ and $\text{L}/b=32$ (for more details see Ref.~\cite{Beane:2012vq}).
Since the volumes are cubic and the two-nucleon systems considered have zero total momentum, the spectra obtained are those of the irreducible representations (irreps) of the octahedral symmetry group \Oh.

The sink operators are defined as products of single nucleon operators in momentum space. The single nucleon operators were designed to have good overlap with the single nucleon ground states~\cite{Basak:2005ir,Basak:2007kj,Morningstar:2013bda}. Let $R$ be an element of \Oh.  Let $N_{m_{s_1}}^{m_{I_1}}(\textbf{k})$ be a nucleon operator with spin and isospin $z$--components $m_{s_1}$ and $m_{I_1}$ and momentum $\textbf{k}$ at time $t$. 
Due to the periodic boundary conditions, the free momenta satisfy $\textbf{k}=2\pi \textbf{n}/L$, where $\textbf{n}$ is an integer triplet. 
With these we can construct two nucleon operators with angular momentum $J m_J$ and isospin $Im_I$
\begin{align}
\label{eq:OpsJlS}
&\mathcal{O}^{Jm_J}_{Im_I;S\ell}(|\textbf{k}|) =
\sum_{\substack{m_S,m_{\ell} \\ m_{s_1},m_{s_2} \\ m_{I_1},m_{I_2}}}
C^{Jm_J}_{\ell m_{\ell},Sm_S}
C^{Sm_S}_{s_2m_{s_2},s_1m_{s_1}}
C^{Im_I}_{\frac{1}{2}m_{I_1},\frac{1}{2}m_{I_2}}
\sum_{R \in \mathcal{O}_h} Y_{\ell m_{\ell}}(\widehat{R \textbf{k}})~ 
N_{m_{s_1}}^{m_{I_1}}({R \textbf{k}}) N_{m_{s_2}}^{m_{I_2}}({ -R \textbf{k}} ),
\end{align}
where $\ell$ and $S$ denote the total orbital angular momentum and spin of the system and $\widehat{R \textbf{k}}$ is the unit vector in the $R \textbf{k}$ direction and the $Y_{\ell m_{\ell}}(\widehat{R \textbf{k}})$ are the standard spherical harmonic functions.
The standard Clebsch-Gordan coefficients, $C^{Jm_J}_{\ell m_{\ell},Sm_S} = \langle J m_j| \ell m_l, S m_s\rangle$ project the operators to total spin S, angular momentum $J$, and isospin $I$.
The infinite volume quantum numbers are not good quantum numbers in a finite volume.
Operators with different angular momentum labels will mix due to the non-spherically symmetric finite spatial boundary conditions. 
To project the operators above to an operator that is in a row $\mu$ of the irrep $\Lambda$ of \Oh, we use the subduction coefficients, $[{C}^{J}_{\Lambda}]_{\mu,m_J}$, found in Ref.~\cite{Dudek:2010wm}
\begin{align}
\mathcal{O}_{\Lambda\mu,Im_I}^{[J\ell S]}(|\textbf{k}|)&=
\sum_{m_J}~[{C}^{J}_{\Lambda}]_{\mu,m_J}~\mathcal{O}_{Jm_JIm_I;S\ell} (|\textbf{k}|).
\end{align}

\begin{figure*}[t]
\centering
\vspace{1cm}
\includegraphics[width=0.99\textwidth]{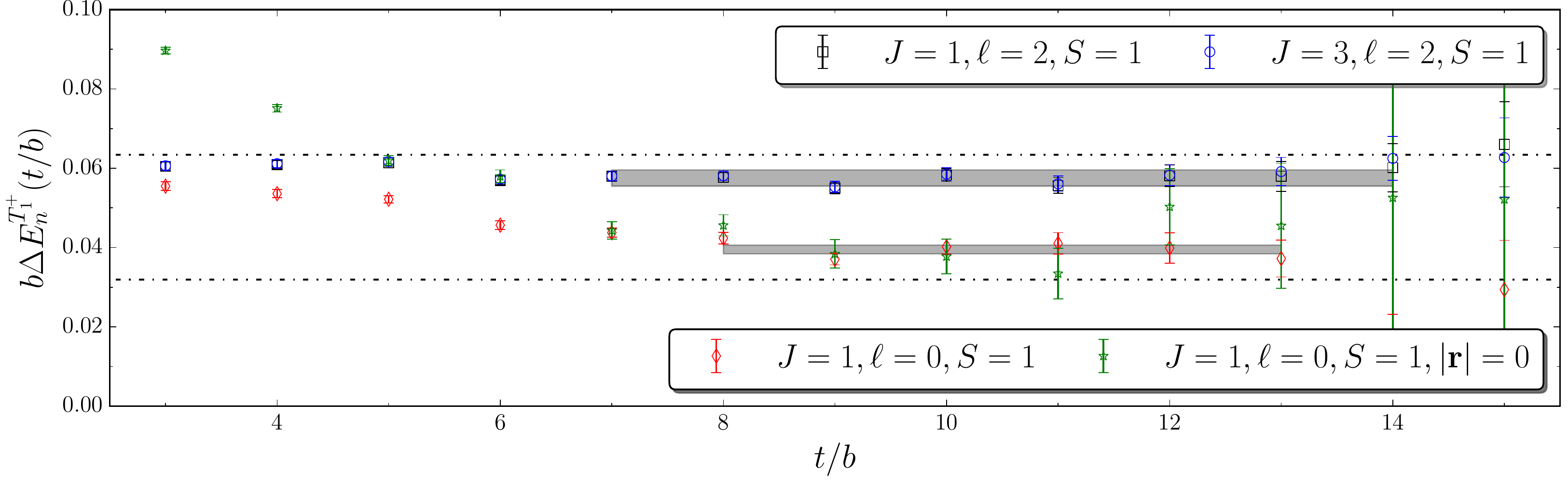}
\caption{\label{fig:diff_waves} Effective masses for the energy splitting, $\Delta E_n= 2\sqrt{m_N^2 +q_n^2}-2m_N$, in lattice units for the second excited state in the spin triplet $T_1^+$ channel at $L/b=32$, showing operators having different $[J\ell S]$ labels (\Eq{OpsJlS}): $J=1,\ell=2,S=1$ (black), $J=3,\ell=2,S=1$ (blue), $J=1,\ell=0,S=1$ (red), $J=1,\ell=0,S=1$, $\mathbf{r} = 0$ (green). The dashed horizontal lines represent the energy levels of the nearest non-interacting two-nucleon states.}
\end{figure*}

While of course it is possible to bypass the construction of operators with $[J \ell S]$ labels in \Eq{OpsJlS} and go directly to a set of operators belonging to definite cubic irreps, it is convenient to keep separate operators of the same cubic irrep having different $[J\ell S]$ labels, as these operators in some cases have very different overlap with the various excited states of a particular cubic irrep. For example, in the spin singlet, $T_1^{+}$ channel, the second non-zero momentum shell of the non-interacting system contains two degenerate states, whose energies split once interactions are turned on. We find that the $T_1^{+}$ operators having $\ell=0$ labels exhibit good overlap with the lower of these two states, while operators with $\ell=2$ labels overlap well onto the higher state (see \Fig{diff_waves}).

Ideally, we would construct NN operators in momentum space at both the source and sink locations, as is done in two-meson calculations~\cite{Dudek:2012gj,Dudek:2012xn}.
However, the computational cost of performing these calculations for two nucleons is orders of magnitude greater than for two mesons.
For this reason, NN calculations typically are performed with local or volume sources for the nucleons at the source and then projected to definite momentum at the sink.

\begin{figure*}[t]
\centering
\vspace{1cm}
\includegraphics[width=0.49\textwidth]{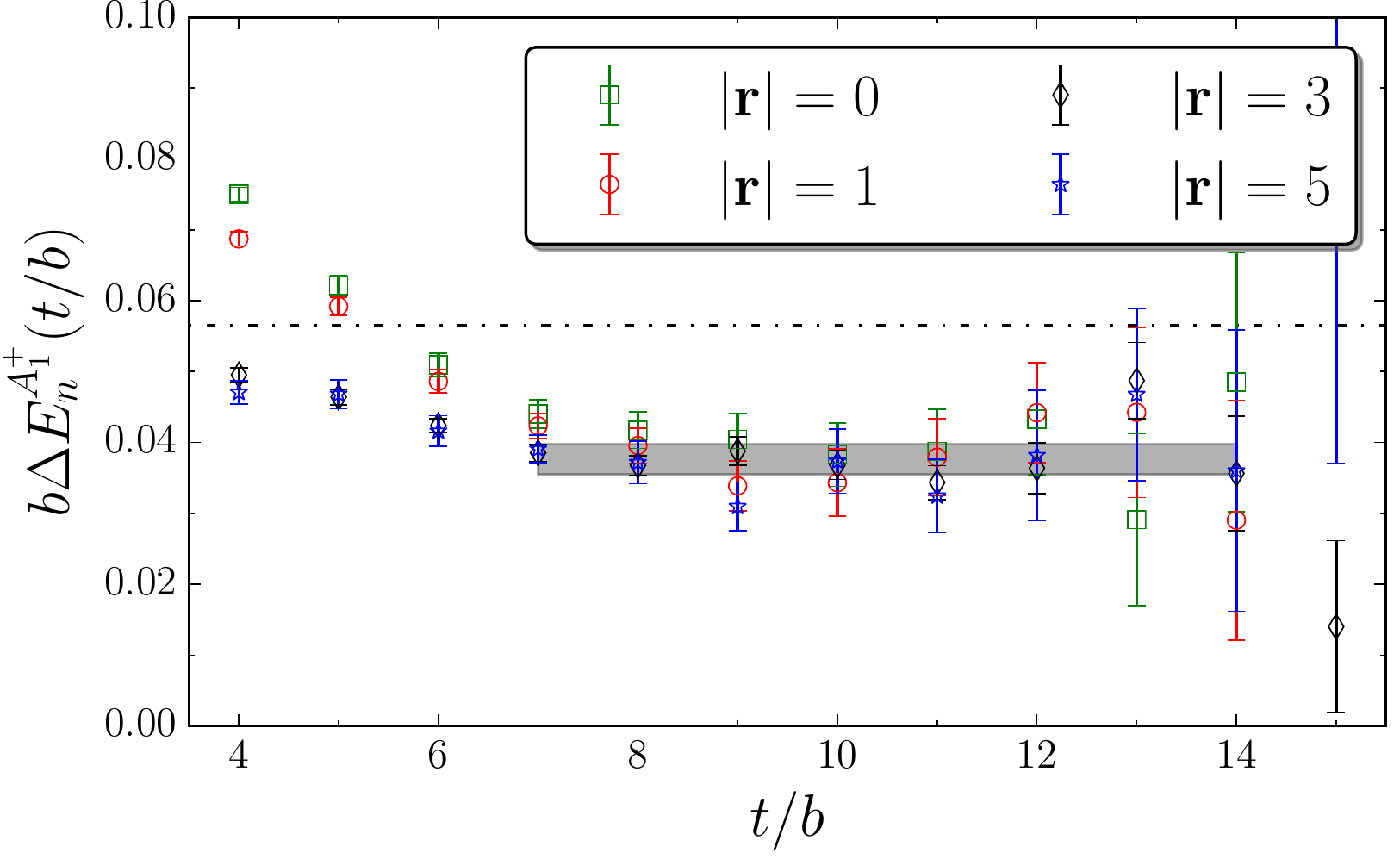}
\includegraphics[width=0.49\textwidth]{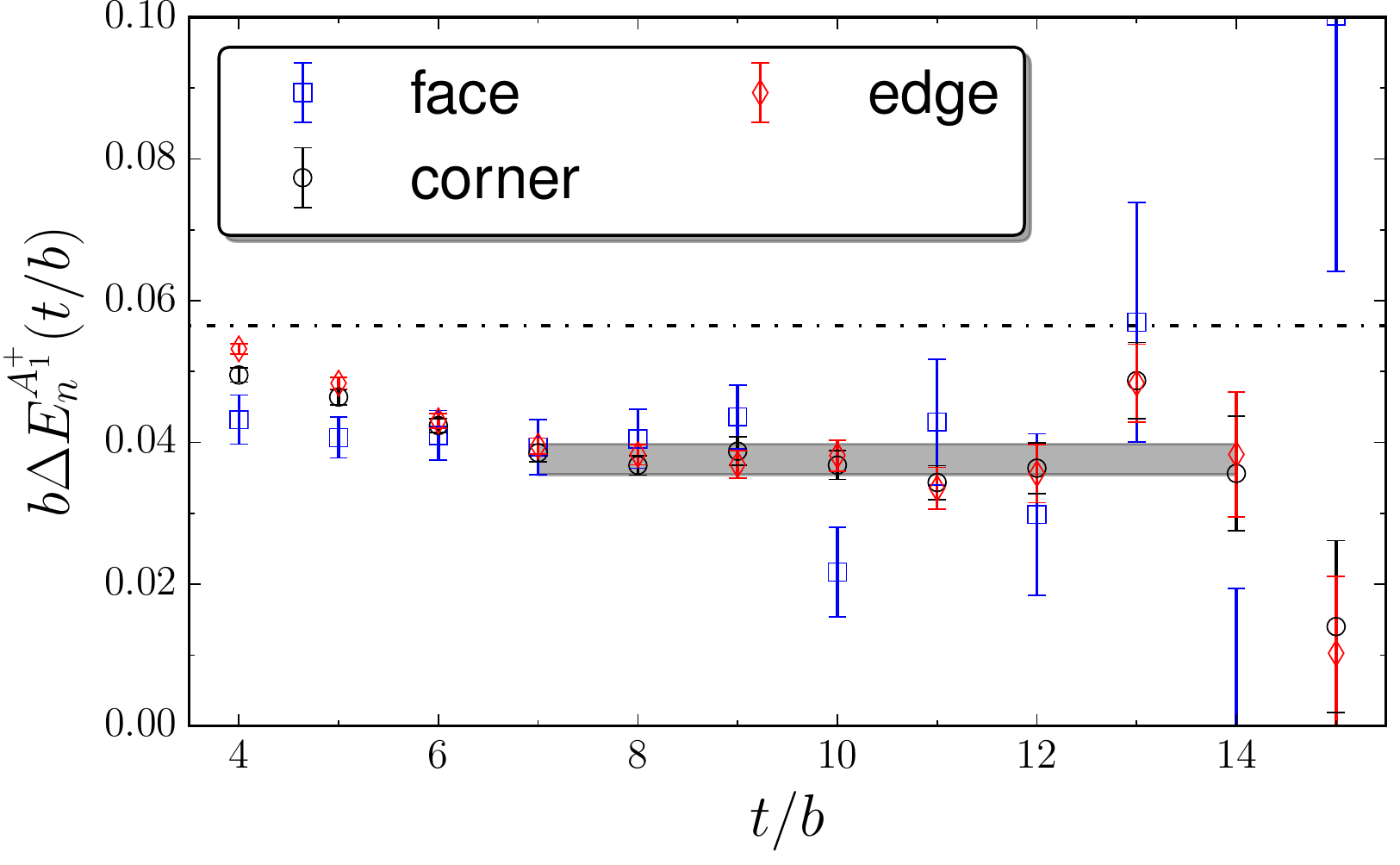}
\caption{\label{fig:disp_source} Effective masses for the energy splitting, $\Delta E_n= 2\sqrt{m_N^2 +q_n^2}-2m_N$, in lattice units for the first excited state in the spin singlet $A_1^+$ channel at $L/b=24$. On the left, we show effective masses corresponding to different displacement lengths at the source: $|\mathbf{r}| = 0$ (green), $|\mathbf{r}| = 1$ (red), $|\mathbf{r}| = 3$ (black), $|\mathbf{r}| = 5$ (blue). On the right, we show effective masses corresponding to different geometric displacements at the source (see text): face (blue), corner (black), edge (red) (``face" sources calculated on a significantly smaller sample of configurations). The dashed lines represent the energy level of the nearest non-interacting two-nucleon state.}
\end{figure*}

The significant improvement made over previous works is the use of spatially displaced nucleon operators at the source, $N^\dagger(t_0,\mathbf{x}_0 + \mathbf{r}/2)N^\dagger(t_0,\mathbf{x}_0 - \mathbf{r}/2)$. 
By displacing the two nucleons at the source, we find a significant increase in the overlap of the operators onto the NN $A_1^+$ and $T_1^+$ irreps ($\sim S$-wave) as compared to the local operators (see \Fig{disp_source}, left).
Further, without such a displacement, for zero total momentum, the overlap of the local operators ($\mathbf{r}=\mathbf{0}$) with the cubic irreps that contain the $P, D$ and $F$ waves are zero, prohibiting a determination of the spectrum in these irreps. 

Displaced operators give us further freedom in designing our sources to have good overlap with the desired states by choosing from various geometries for the displacements. For example, after fixing one nucleus to a single lattice point, $(0,0,0)$, one may then calculate the set of pairs where the second nucleon is displaced in all possible ways along a single axis, $\mathbf{r} = |\mathbf{r}| (0,0,1)$ (plus all cubic rotations), or along multiple axes, such as $\mathbf{r} = |\mathbf{r}| (0,1,1)$ (plus all cubic rotations) and $\mathbf{r} = |\mathbf{r}| (1,1,1)$ (plus all cubic rotations). We have named these geometries ``face", ``edge", and ``corner", respectively, and disregard more complicated geometries. Each collection of geometries may then be projected onto the desired cubic irrep as described above for the sink operators (\Eq{OpsJlS}), with the momentum vectors $\mathbf{k}$ replaced by the set of displacements vectors, $\mathbf{r}$. Note, however, that this is only a partial projection because we use a reduced set of displacement vectors compared to the full lattice volume. Example effective masses for the three types of geometries are shown in \Fig{disp_source} (right). While the ``face" sources have a reduced computational cost compared to ``corner" and ``edge" sources (7 inversions versus 9 (``corner") and 13 (``edge")), they have zero overlap with several channels of interest due to their simple geometrical structure. We have chosen to focus on ``corner" sources for the remainder of this work, to balance good overlap with a large number of states with moderate computational cost.

\begin{figure*}[t]
\centering
\vspace{1cm}
\includegraphics[width=0.49\textwidth]{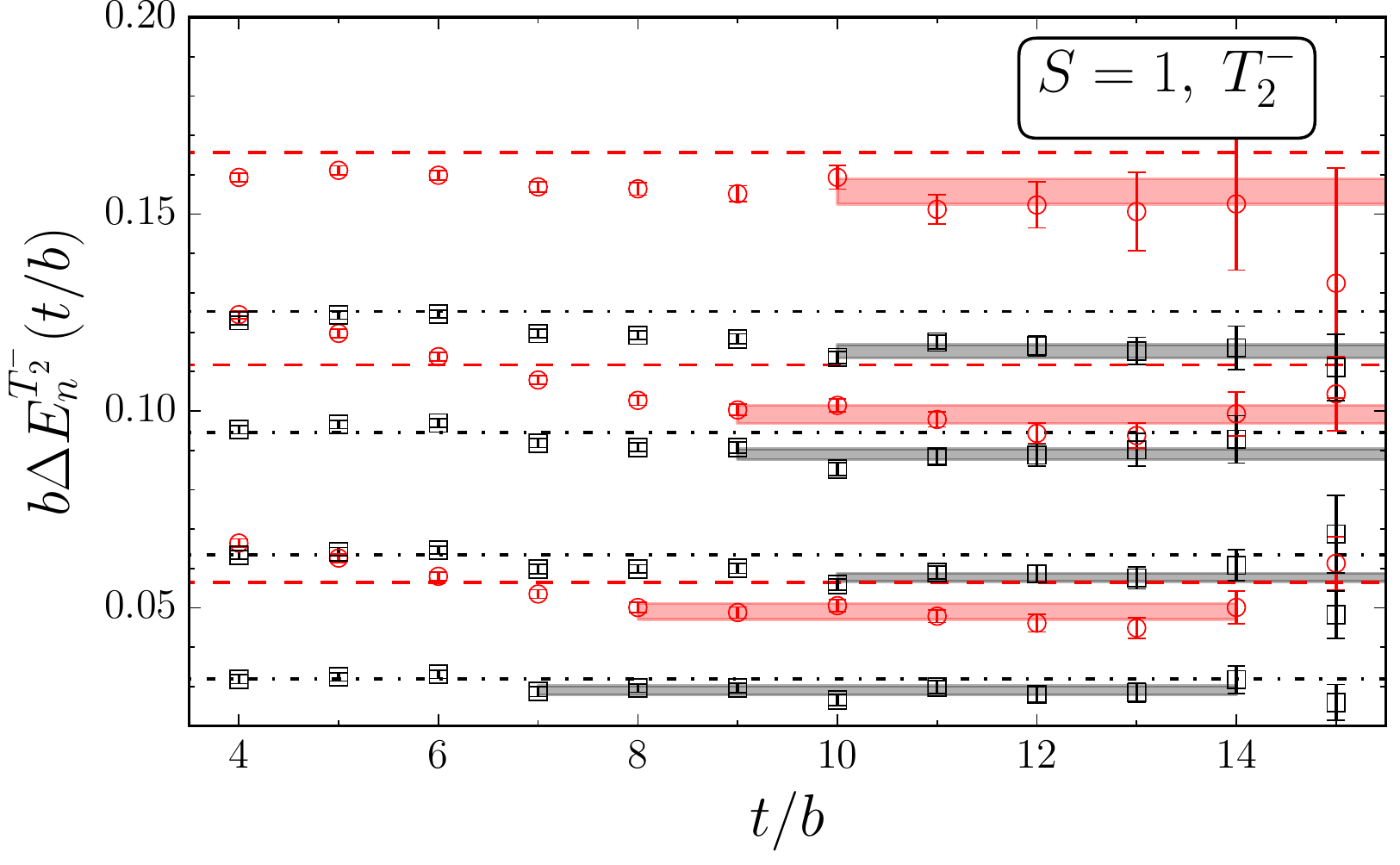}
\includegraphics[width=0.49\textwidth]{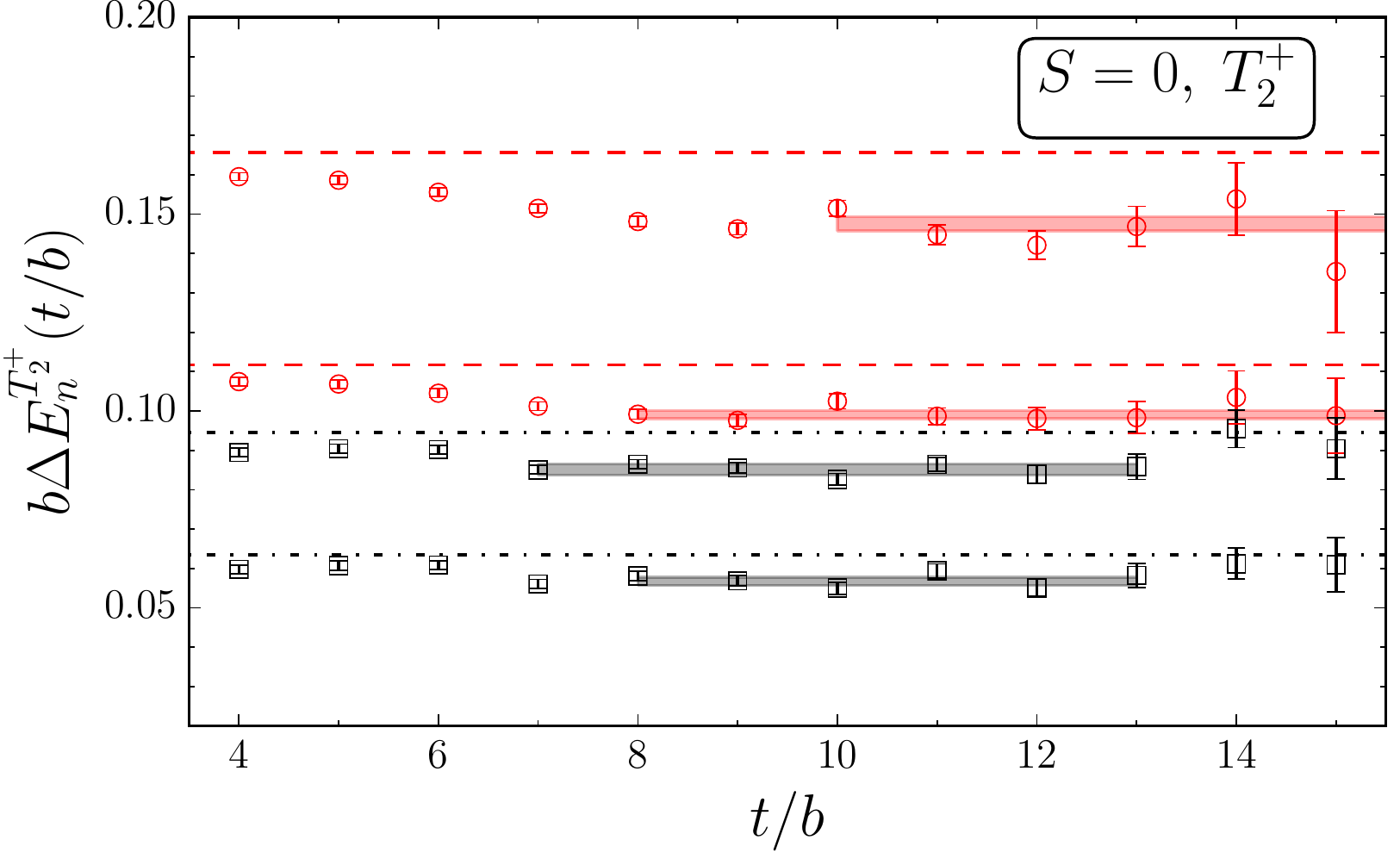}
\caption{\label{fig:effective_masses} Effective masses for the energy splittings, $\Delta E_n= 2\sqrt{m_N^2 +q_n^2}-2m_N$, in lattice units for spin triplet $T_2^-$ and spin singlet $T_2^+$ channels. Several energy levels are shown (corresponding to the non-interacting shell labeled by $n$), and red circles (black squares) points correspond to $L/b=24$ ($32$). Horizontal bands represent fits to the data (see text). Dashed (dot-dashed) lines represent the energy levels of the non-interacting two-nucleon systems.}
\end{figure*}

In Fig.~\ref{fig:effective_masses}, we show effective mass plots for several operators in two different cubic irreps along with the resulting determination of the energy levels. We construct all sink operators that have free momenta $|\textbf{k}|L/2\pi\leq \sqrt{6}$. 
Note that the $T_2^-$ channel has overlap with the physical $^3P_2,\,^3F_2$ and $^3F_4$ channels, and the $T^+_2$ channel has overlap with the physical $^1D_2$ channel. Both channels have additional overlap with kinematically suppressed higher partial waves.
The bands denote the combined statistical and systematic uncertainties obtained by performing 
one and two exponential fits to the correlation functions over a range of time windows.
The fit ranges displayed are representative of the times considered in the fits.
For reference, the dashed lines correspond to the energy levels of a non-interacting two-nucleon system.
The calculations were performed with 20 sources on 3822 configurations and 70 sources on 1018 configurations for the $L/b=24$ and $L/b=32$ volumes respectively.

\section{NN scattering phase shifts \label{sec:results}}
In general, due to the reduction of rotational symmetry in a finite volume, there is not a one-to-one correspondence between the finite volume spectrum and the infinite volume scattering amplitudes. For sufficiently low energies, one can expect higher partial waves to be kinematically suppressed.  
Ignoring partial wave mixing and $\ell\geq4$ waves, the spectra of a number of cubic irreps satisfy the quantization condition~\cite{Briceno:2013lba}
\begin{eqnarray}
\label{eq:1D_QCs}
\frac{q\cot\delta_\Lambda(q)}{4\pi}= c_{00}(q^{2})
	+\alpha_{4,\Lambda} \frac{c_{40}(q^{2})}{q^4}
	+\alpha_{6,\Lambda} \frac{c_{60}(q^{2})}{q^6},
\end{eqnarray}
where $q$ is the on-shell relative momentum of the system, $q^{2}=\frac{E^{2}_{NN}}{4}-m_{N}^2$, $\delta_\Lambda(q)$ is the scattering phase shift of the partial wave that primarily couples to the $\Lambda$ irrep, $\alpha_{\ell,\Lambda}$ are constants, Ref.~\cite{Briceno:2013lba},
and the $c_{\ell m_\ell}$ are kinematic, non-linear functions that depend solely on the momentum and the volume 
\begin{eqnarray}\label{eq:clm}
c_{\ell m_\ell }(q^{2})
&=&\frac{\sqrt{4\pi}}{L^3}\left(\frac{2\pi}{L}\right)^{\ell -2}~\sum_{\mathbf r \in \mathbb{Z}^3}\frac{|\mathbf{r}|^\ell Y_{\ell m_\ell }(\mathbf{r})}{(r^2-q^2)}. \label{eq:clm}
\end{eqnarray}

\begin{figure*}[t]
\begin{tabular}{ccc}
\centering
\includegraphics[scale=0.3]{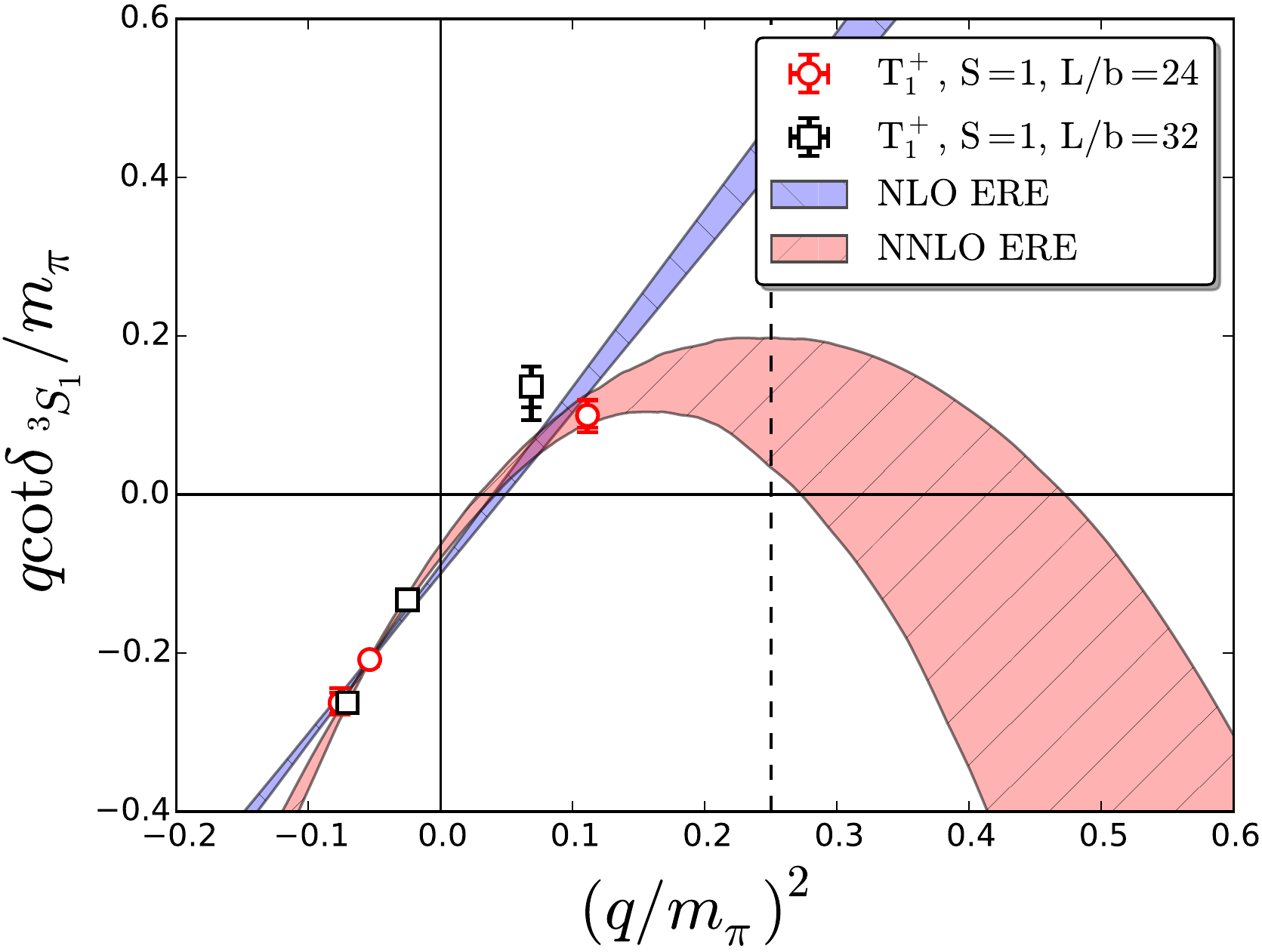}&
\hspace{.8cm}\includegraphics[scale=0.3]{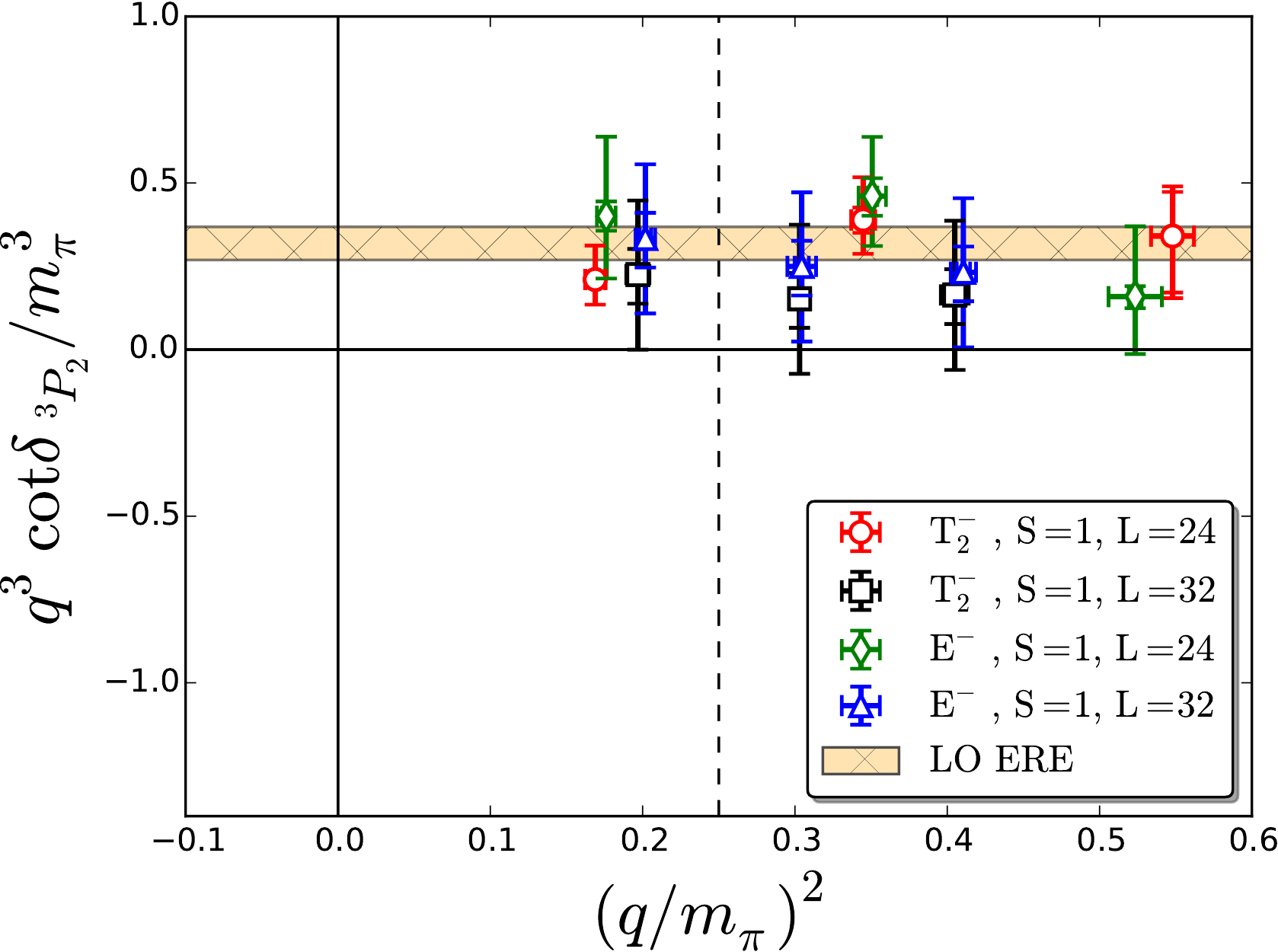}&
\hspace{.95cm}\includegraphics[scale=0.3]{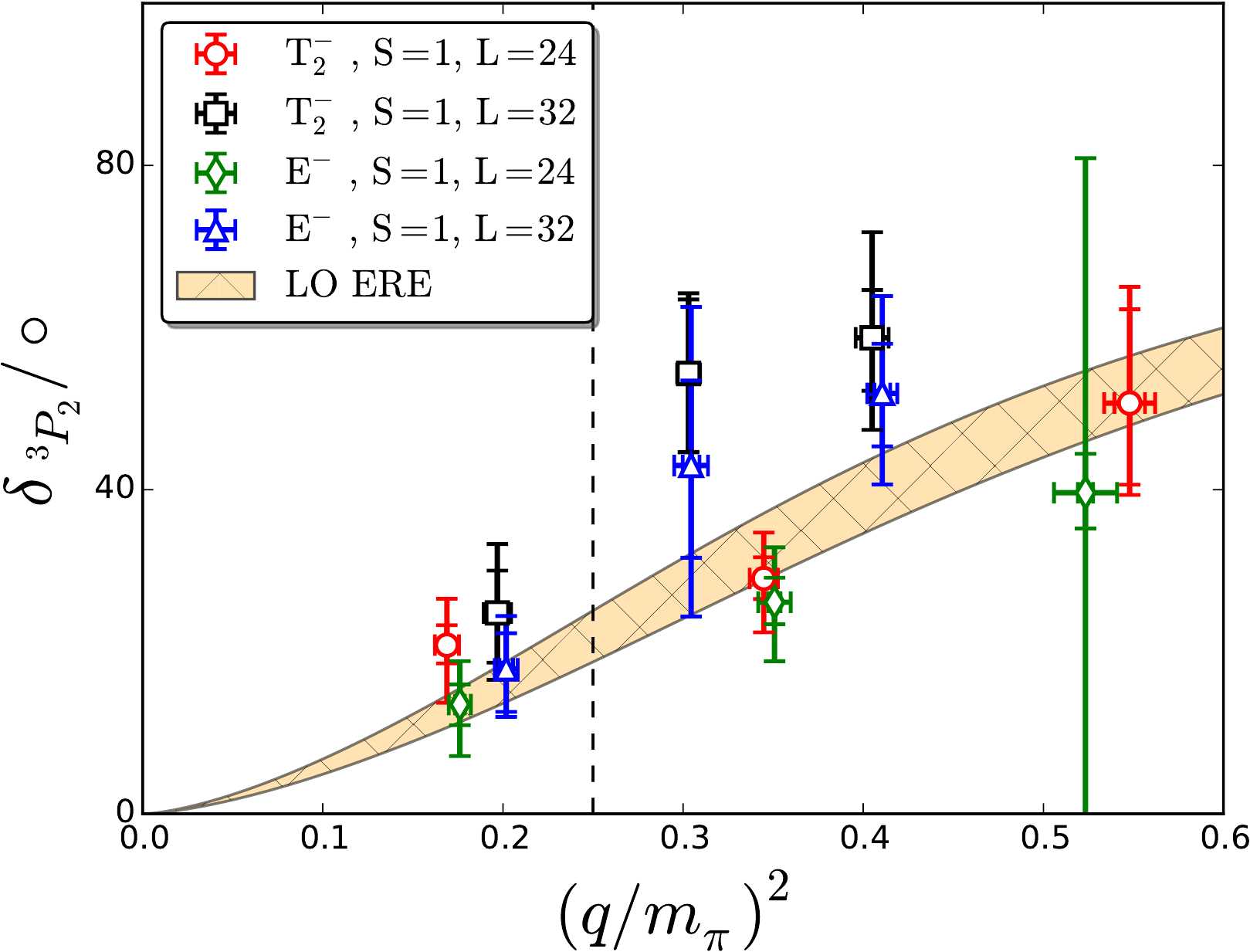}
\\
\includegraphics[scale=0.3]{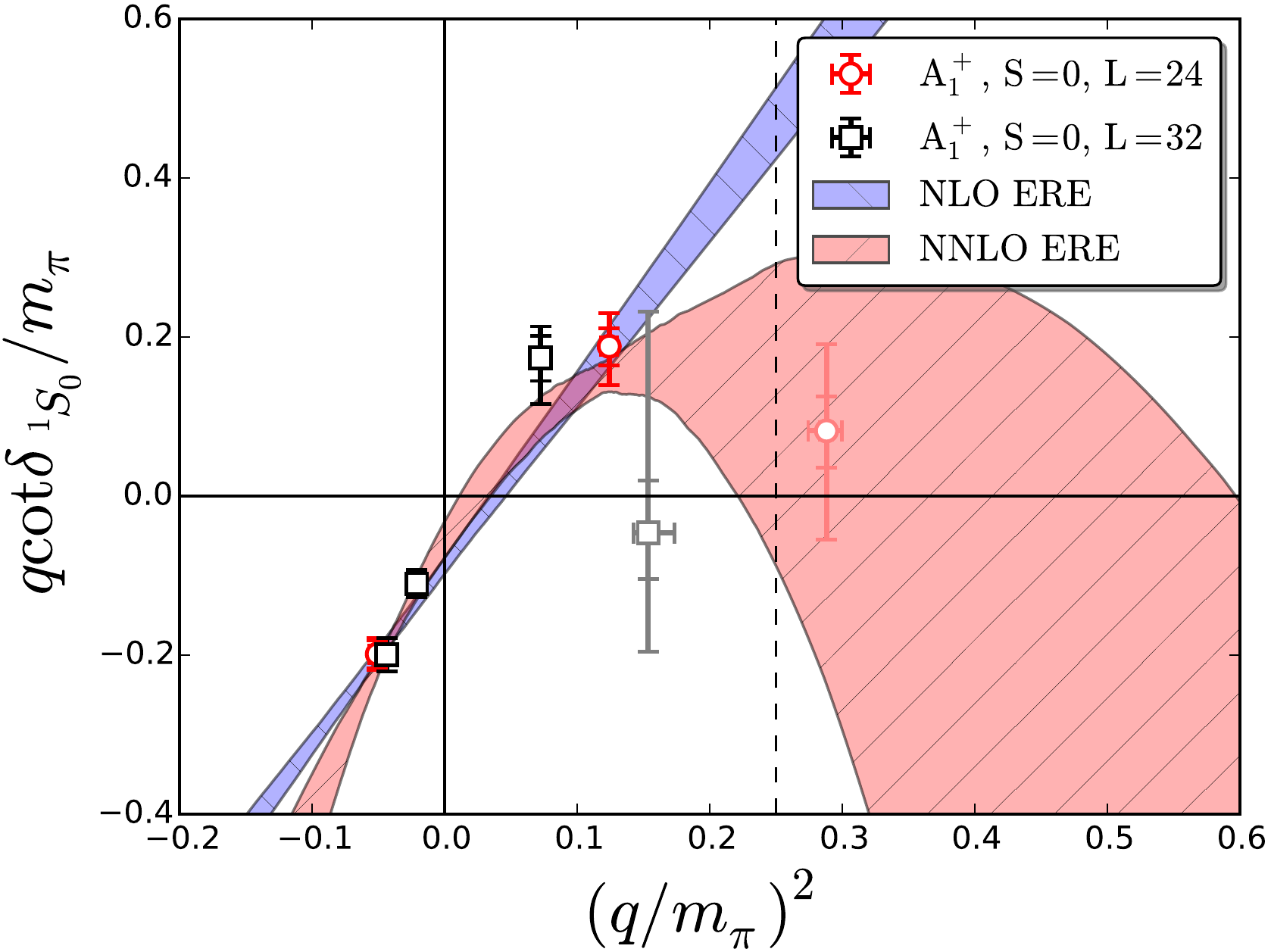}&
\hspace{.8cm}\includegraphics[scale=0.3]{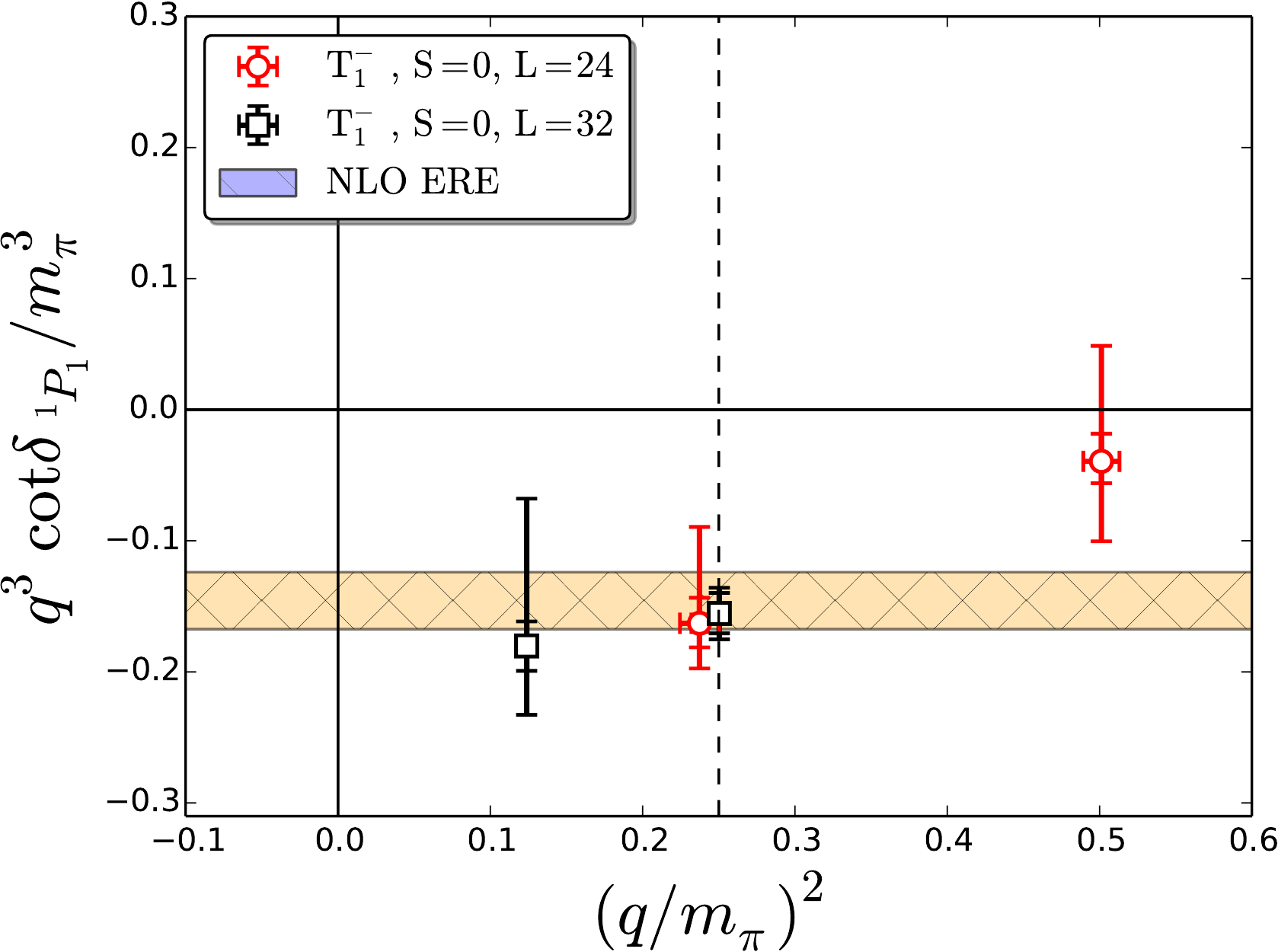}&
\hspace{.8cm}\includegraphics[scale=0.3]{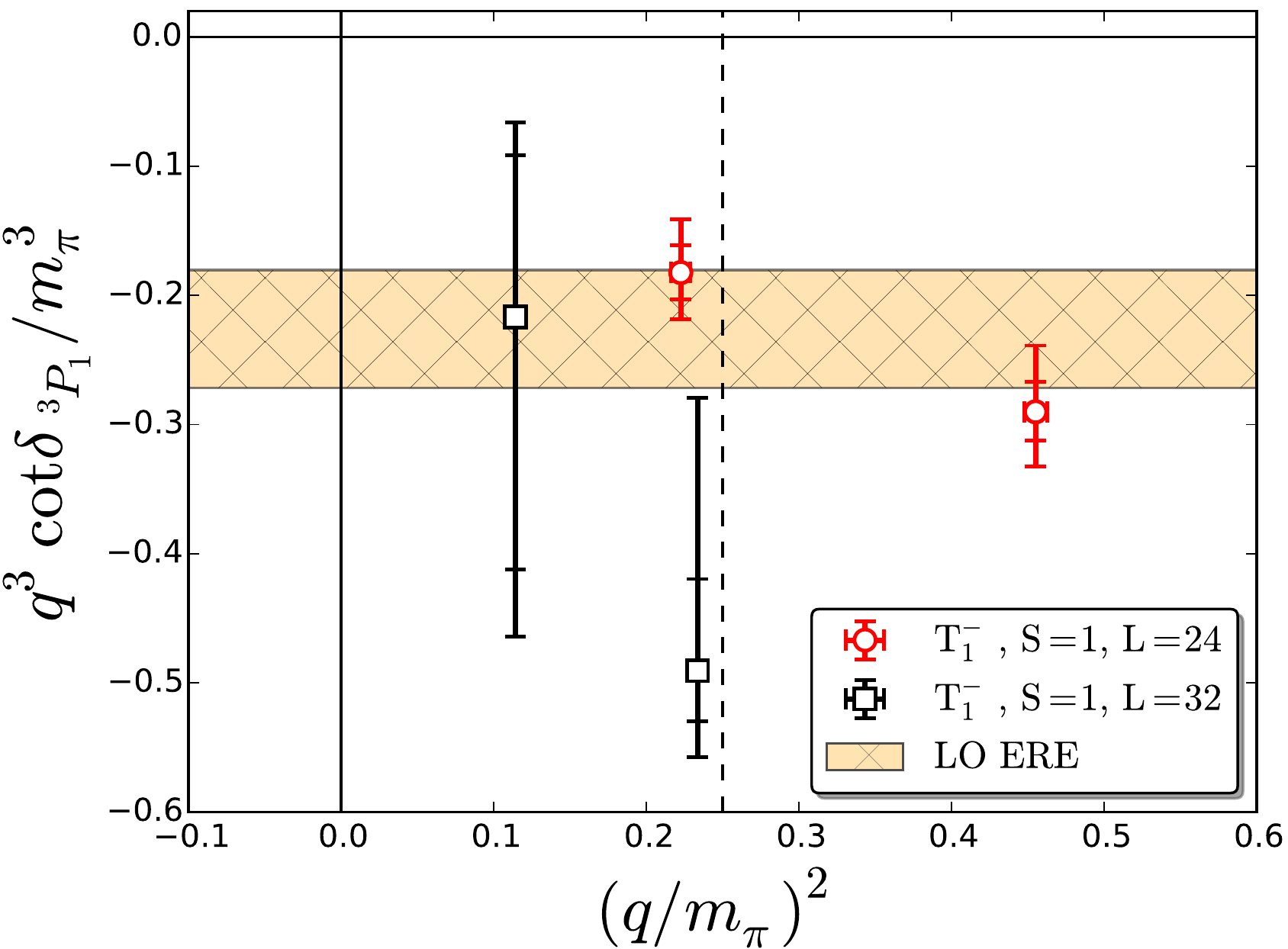}
\\
\includegraphics[scale=0.3]{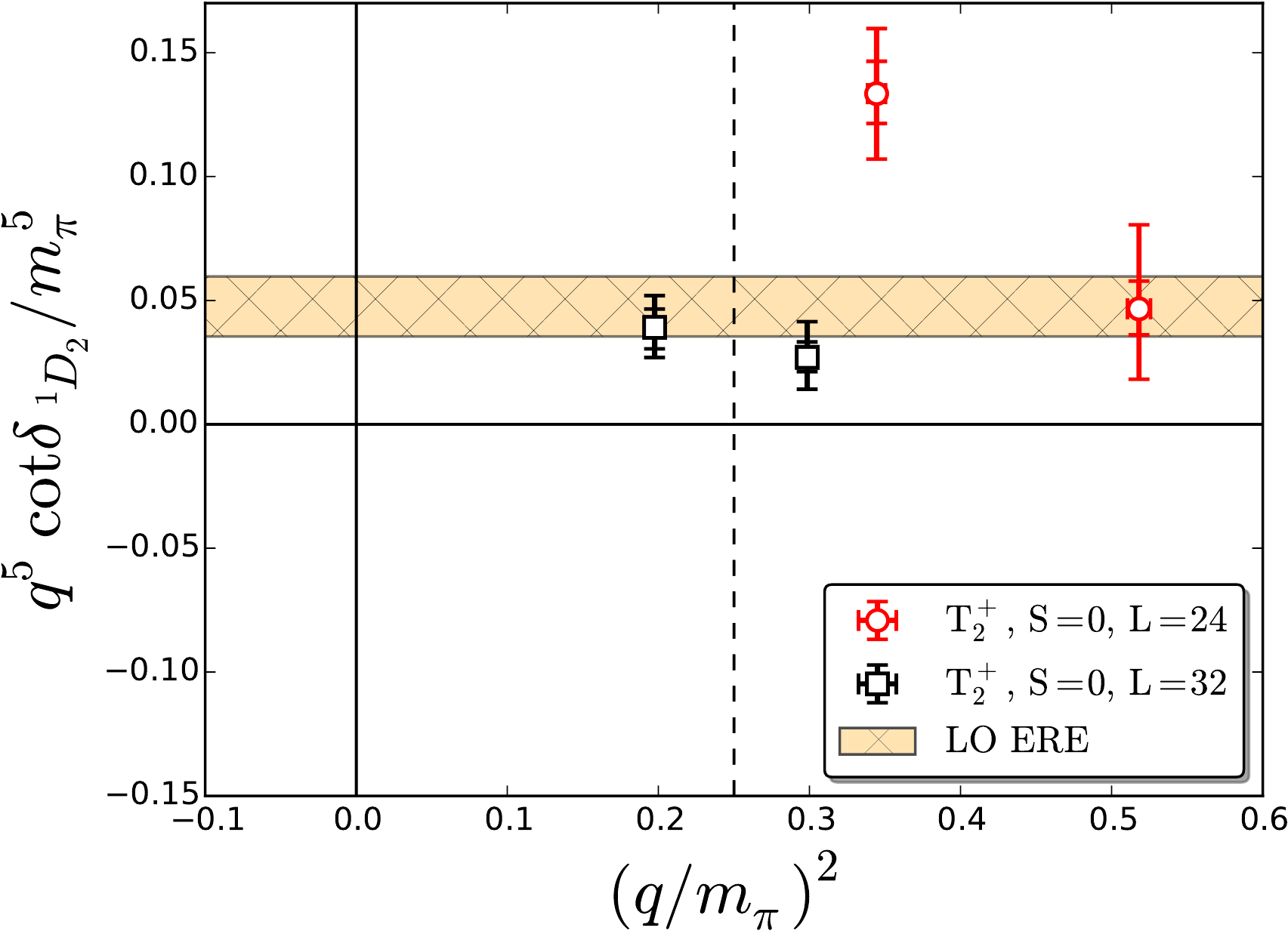}&
\hspace{.8cm}\includegraphics[scale=0.3]{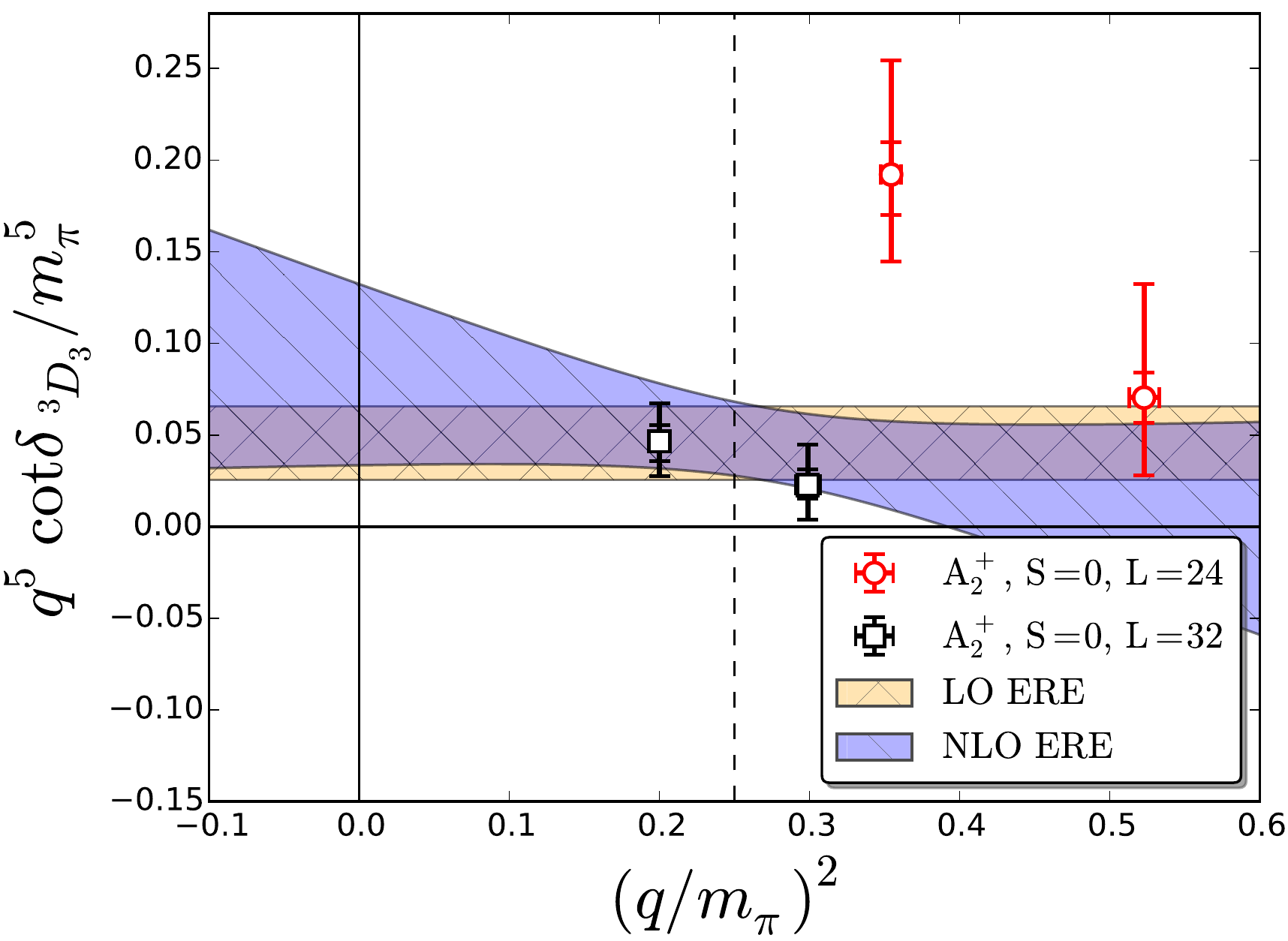}&
\hspace{.8cm}\includegraphics[scale=0.3]{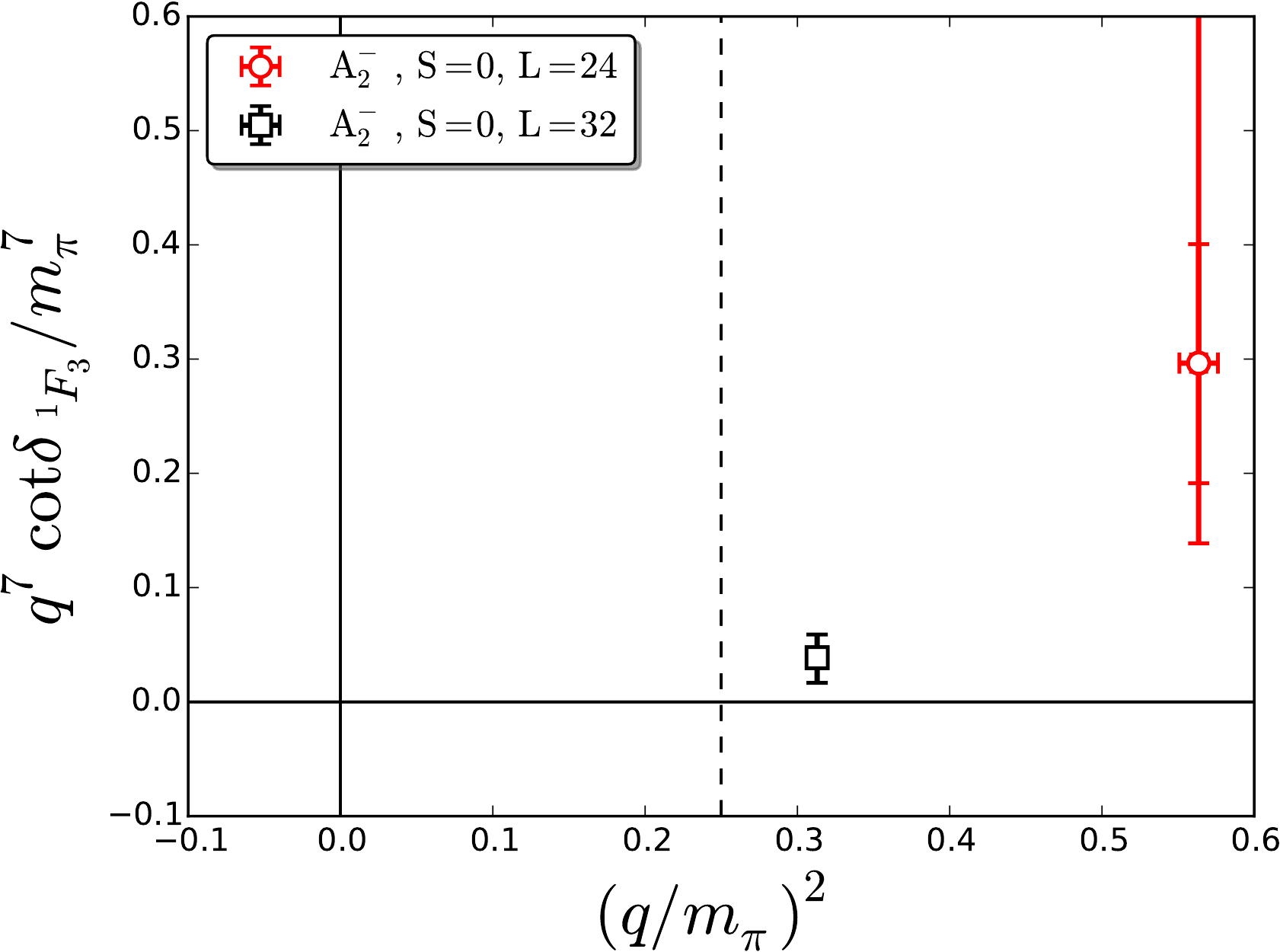}
\end{tabular}
\caption{\label{fig:Apipi_to_pi_abs_triplet}
Shown are examples of the phase shift determination in several partial wave channels, as well as representative ERE fits. The dashed vertical lines indicate the momentum at which the $t$-channel cut occurs ($q=m_{\pi}/2$). In most panels we plot $q^{2\ell+1}\cot\delta_\ell$ which is used to determine the parameters in the ERE. In the $^1S_0$ channel, the faded points were not included in the ERE fit, but are displayed here to show consistency. In the upper right panel we also show the phase shift $\delta_{\,^3\!P_2}$ as a function of the lattice momenta.
}
\end{figure*}

Employing Eqs.~(\ref{eq:1D_QCs}) and (\ref{eq:clm}), we obtain the scattering phase shifts evaluated at the on-shell relative momenta derived from the spectrum. In Fig.~\ref{fig:Apipi_to_pi_abs_triplet}
we give illustrative examples of the quality of the results for the spin-triplet and spin-singlet channels in various irreps that couple predominantly to a given partial wave. The bands are fits to the effective range expansion (ERE) to different orders, i.e.
\begin{equation}
q^{2\ell+1}\cot\delta_\ell = -\frac{1}{a_\ell}+\frac{1}{2}r_\ell\,q^2
	+\frac{1}{4!}P_\ell q^{4}+\mathcal{O}(q^6),
\label{eq:ERE}
\end{equation} 
where $a_\ell$, $r_\ell$, and $P_\ell$ are the scattering length, effective range, and shape parameter for $\ell=0$ and the corresponding parameters of the ERE for $\ell>0$.
Results for LO ($q^0$) fits are denoted by yellow bands, fits to NLO ($q^2$) are blue, and NNLO ($q^4$) fits are red. The ERE parameters determined from these fits are listed in Table~\ref{tab:ere_params}. The dashed vertical line represents the t-channel cut, above which the ERE is expected to break down. 
While the ERE becomes formally unjustifiable past this cut, the L\"uscher formalism holds for all energies below the $NN\pi$ threshold, well above the energies considered.

We obtain results significantly different from zero for various channels, including $P$ and $D$ channels.
One beautiful illustration of the power of the operators and the finite volume formalism being used is that of ${\,^3\!P_2}$ in Fig.~\ref{fig:Apipi_to_pi_abs_triplet} (top-middle and top-right).
In this channel the spectrum was determined with two irreps that have overlap with the same partial wave using two different volumes.
The consistency of the extracted phase shifts clearly demonstrates that the generalization of the L\"uscher formalism for NN-systems is working~\cite{Briceno:2013lba}.
We find that over the kinematic range of our calculations, $q^{3}\cot\delta_{\,^3\!P_2}$ is consistent with a constant, even above the t-channel cut.  
Furthemore, we find no evidence of the t-channel cut playing an important role for any of the channels studied at these values of the quark masses.

Currently, we ignore partial wave mixing, an issue that will be addressed in subsequent publications. This formidable challenge has only been addressed in two-meson calculations~\cite{Dudek:2012xn,Dudek:2012gj, Wilson:2014cna, Dudek:2014qha, Wilson:2015dqa,Bali:2015gji}. However, some evidence that the mixing from higher partial waves is small in at least one channel can be obtained by investigating the results for the $^3P_2$ channel in Fig.~\ref{fig:Apipi_to_pi_abs_triplet}. Again, for this channel we have two cubic irreps, $T_2^-$ and $E^-$, for which the lowest partial wave is $^3P_2$. Any differences between the two irreps must arise from mixing with higher partial waves. We find that the two cubic irreps give completely consistent results, indicating that this mixing must be too small to resolve within our error bars. Additionally, the first contribution from mixing in the $T_1^-$ cubic irrep, having lowest partial wave $^1P_1$, comes from the $^1F_3$ partial wave. We obtain information on the strength of the $^1F_3$ phase shift independently using the $A_2^-$ cubic irrep, and find it to be extremely small, thus it likely does not contribute to mixing in the $T_1^-$ cubic irrep.

\begin{table}
\begin{ruledtabular}
\begin{tabular}{c|ccc|c||c|c|c}
channel& $1/\left(a_\ell\, m_\pi^{2\ell+1}\right)$& $r_\ell\, m_\pi^{2\ell+1}$& $P_\ell\, m_\pi^{2\ell+3}$& ERE order
&channel& $1/\left(a_\ell\, m_\pi^{2\ell+1}\right)$& ERE order \\
\hline
${}^1 S_0$& \phantom{-}0.088(10)& \phantom{-}4.47(31)& --& $q^2$
&${}^3 P_0$& \phantom{-}0.234(75)& $q^0$\\
${}^1 S_0$& \phantom{-}0.056(24)& \phantom{-}5.45(76)& -202(114)& $q^4$
& ${}^3 P_1$& \phantom{-}0.237(92)& $q^0$\\
${}^3 S_1$& \phantom{-}0.094(06)& \phantom{-}4.27(21)& --& $q^2$
&${}^3 P_2$& -0.317(49)& $q^0$\\
${}^3 S_1$& \phantom{-}0.071(09)& \phantom{-}4.51(24)& -139(38)& $q^4$
& ${}^1 P_1$& \phantom{-}0.146(22)& $q^0$\\
${}^3 D_3$& -0.046(20)& --& --& $q^0$
&${}^1 D_2$& -0.047(12)& $q^0$\\
${}^3 D_3$& -0.082(51)& -0.27(34)& --& $q^2$
\end{tabular}
\end{ruledtabular}
\caption{\label{tab:ere_params}
Effective range parameters for the various scattering channels determined from fits to the orders indicated in the last column.}
\end{table}

\subsection{Bound states from the Effective Range Expansion}

Assuming the results are within the range of convergence of the ERE, the infinite volume bound state energies may be determined by solving for poles in the derived scattering amplitude.
For a single channel scattering amplitude, these satisfy
\begin{eqnarray}\label{eq:q_B}
	q_B\cot\delta(q_B)\equiv iq_B\, ,
\end{eqnarray}
which has a solution below threshold for $q_B=i\kappa_B$. 
In both $S$-wave channels, the ERE expansion appears to be converging well (top-left and middle-left figures in Fig.~\ref{fig:Apipi_to_pi_abs_triplet}) with small $\mathcal{O}(q^4)$ corrections where the results exist below the t-channel cut.
Using this method and the spectrum obtained in this work, we find deeply bound states in both the $^1S_0$ and $^3S_1$ channels, with binding energies of 
\begin{align}
&B_{^1S_0} = 21.8({}_{-5.1}^{+3.2})({}_{-2.8}^{+0.8}) \textrm{ MeV},&
&B_{^3S_1} = 30.7({}_{-2.5}^{+2.4})({}_{-1.6}^{+0.5}) \textrm{ MeV}\, .&
\end{align}
The first uncertainties are our fitting statistical and systematic uncertainties combined in quadrature and the second is an estimate of systematic uncertainties arising from the truncation of the ERE. In addition, in the $^3S_1$ channel we find a second pole near threshold, corresponding to 
\begin{equation}
B_{^3S_1}^{(2)}=3.3({}_{0.9}^{1.0})({}_{0.2}^{0.6}) \textrm{ MeV}\, .
\end{equation}

Corresponding to each of these poles we find finite volume states whose energies are consistent with the expected exponential volume dependence associated with bound states. However, with only two volumes we cannot definitively state whether the volume dependence is exponential or polynomial. 
With this precision, it is unclear whether this shallow bound state corresponds to a true bound state or a near-threshold scattering state.
Improved analysis techniques, such as a full basis of interpolating fields in momentum space, as has been successfully used in the two-meson systems~\cite{Dudek:2012xn, Dudek:2012gj, Wilson:2014cna, Dudek:2014qha,Wilson:2015dqa}, or additional statistics may be necessary to fully settle this matter. We do note, however, that all of the negative energy finite-volume shifts are larger by 4 or more standard deviations than the corresponding finite-volume energy shift for a bound state at threshold, $\Delta E \sim -\frac{3.786}{ML^2}$ \cite{Beane:2003da,Lee:2005fk}, as shown in Table~\ref{tab:unitary_comp}. This indicates that the interactions producing these states are more attractive than those at unitarity, and the states should therefore correspond to true bound states in the infinite volume limit.
Further evidence of multiple bound states are presented in the next section.

\begin{table}
\begin{ruledtabular}
\begin{tabular}{cc|cccc}
$L$& unitarity bound& 
	\multicolumn{2}{c}{$T_1^+ ({}^3 S_1)$}& \multicolumn{2}{c}{$A_1^+ ({}^1 S_0)$}
\\ 
	&$-3.786 / (ML^2)$& $\Delta E_0$& $\Delta E_1$& $\Delta E_0$& $\Delta E_1$\\
\hline
24& -7.41 
	& -30.4(2.4)(5.1)& -21.4(1.0)(0.5)
	& -20.2(2.1)(1.5)& --\\
32& -4.17
	& -28.1(1.8)(2.4)& -9.95(.99)(.42)
	& -17.3(1.7)(2.3)& -8.35(.99)(.48)\\
\end{tabular}
\end{ruledtabular}
\caption{\label{tab:unitary_comp}
Negative shifted energy states in ${}^3 S_1$ and ${}^1 S_0$ channels compared with the unitarity bound.
These energy levels have been converted to MeV using the lattice spacing $b=0.145$~fm.  No scale setting uncertainty is assigned as we are just comparing the energy levels to the unitarity bound.}
\end{table}

\subsection{Evidence for multiple negatively shifted energy states}

\begin{figure*}[t]
\centering
\vspace{1cm}
\includegraphics[scale=0.49]{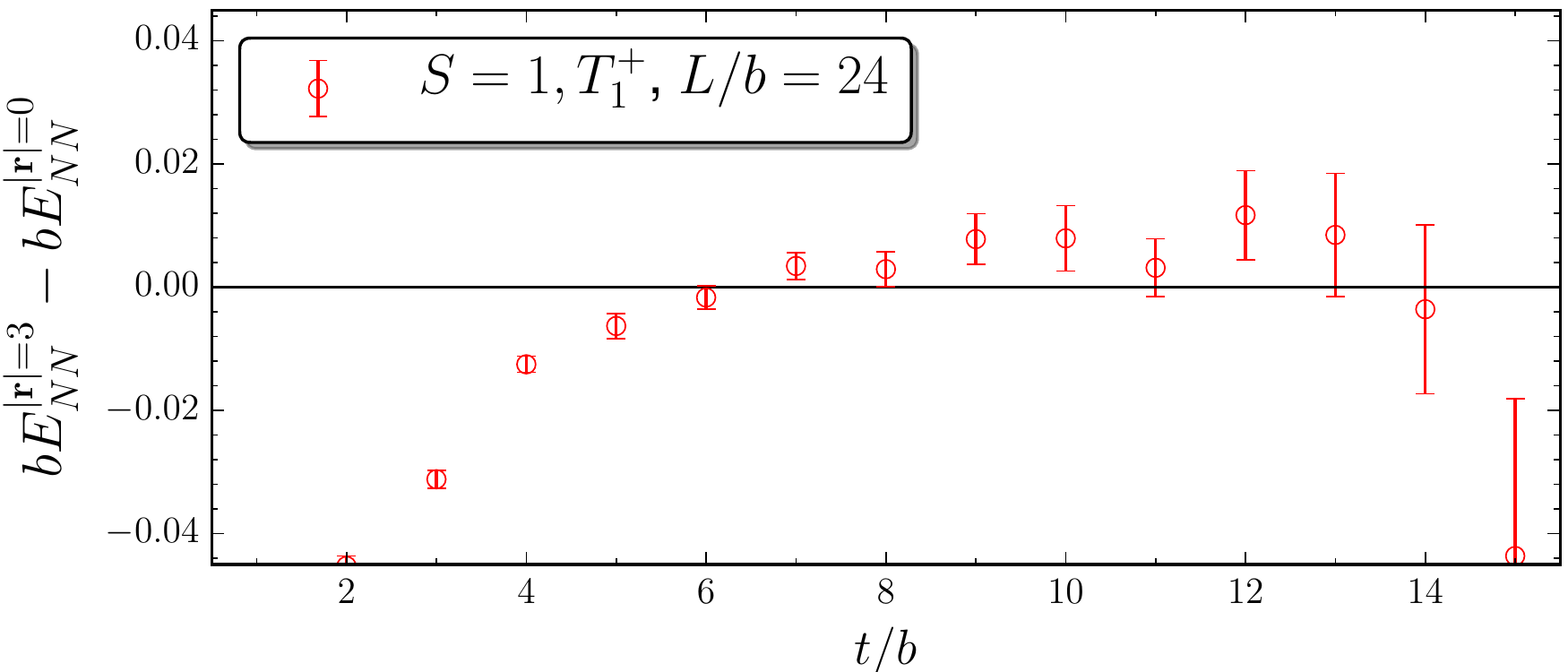}
\includegraphics[scale=0.49]{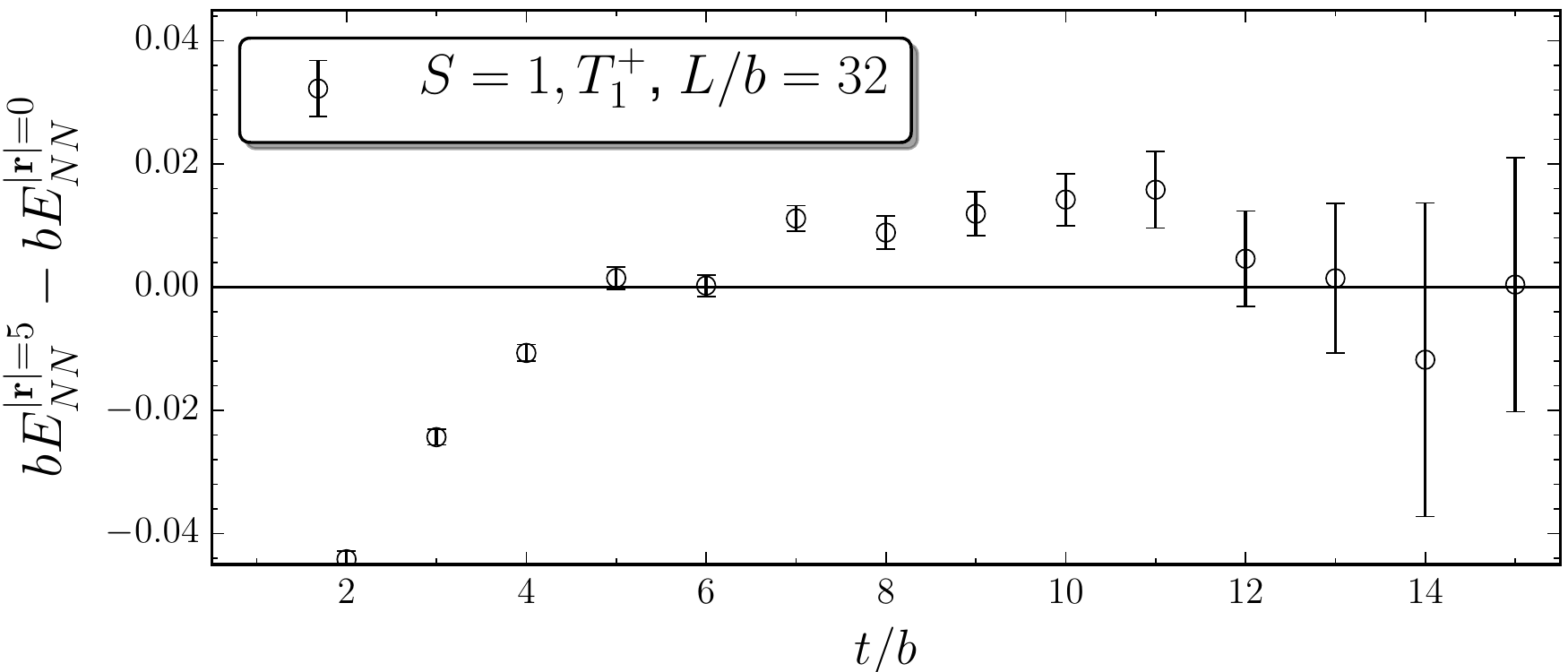}
\includegraphics[scale=0.49]{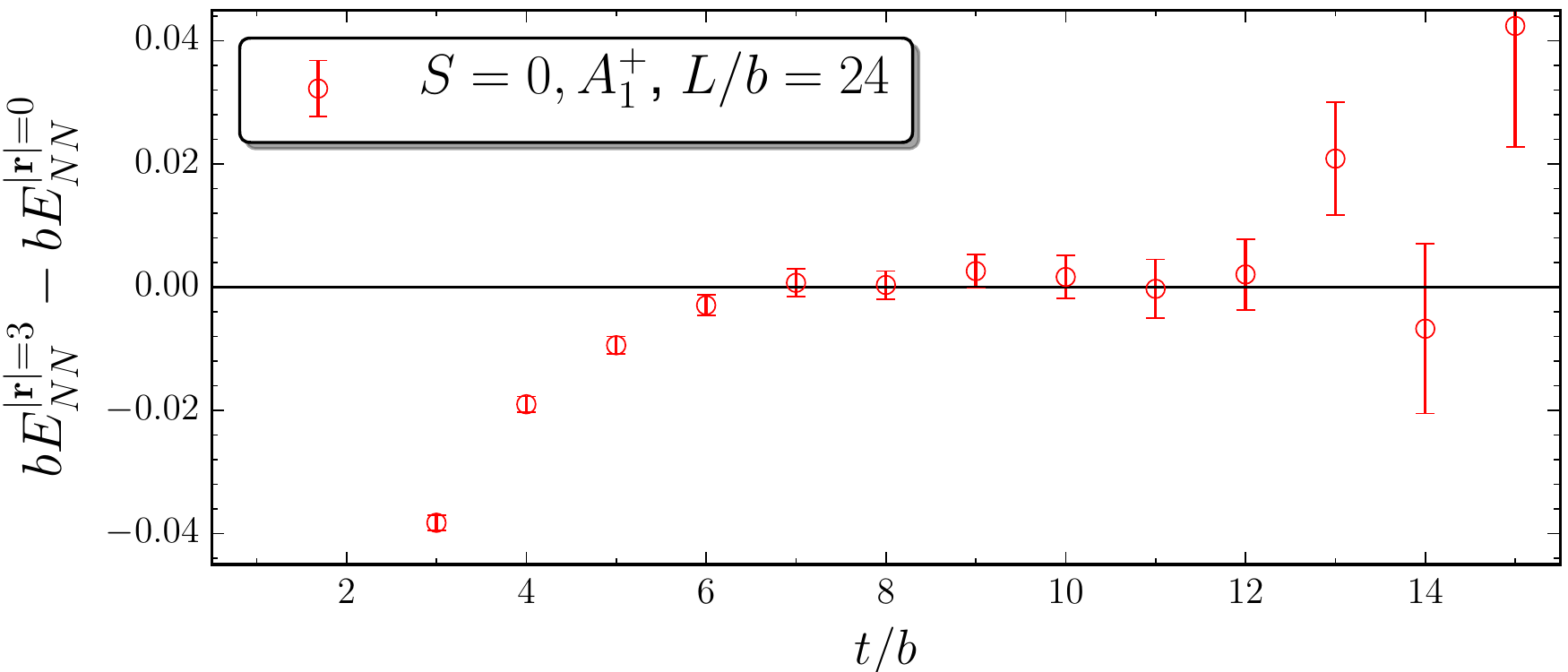}
\includegraphics[scale=0.49]{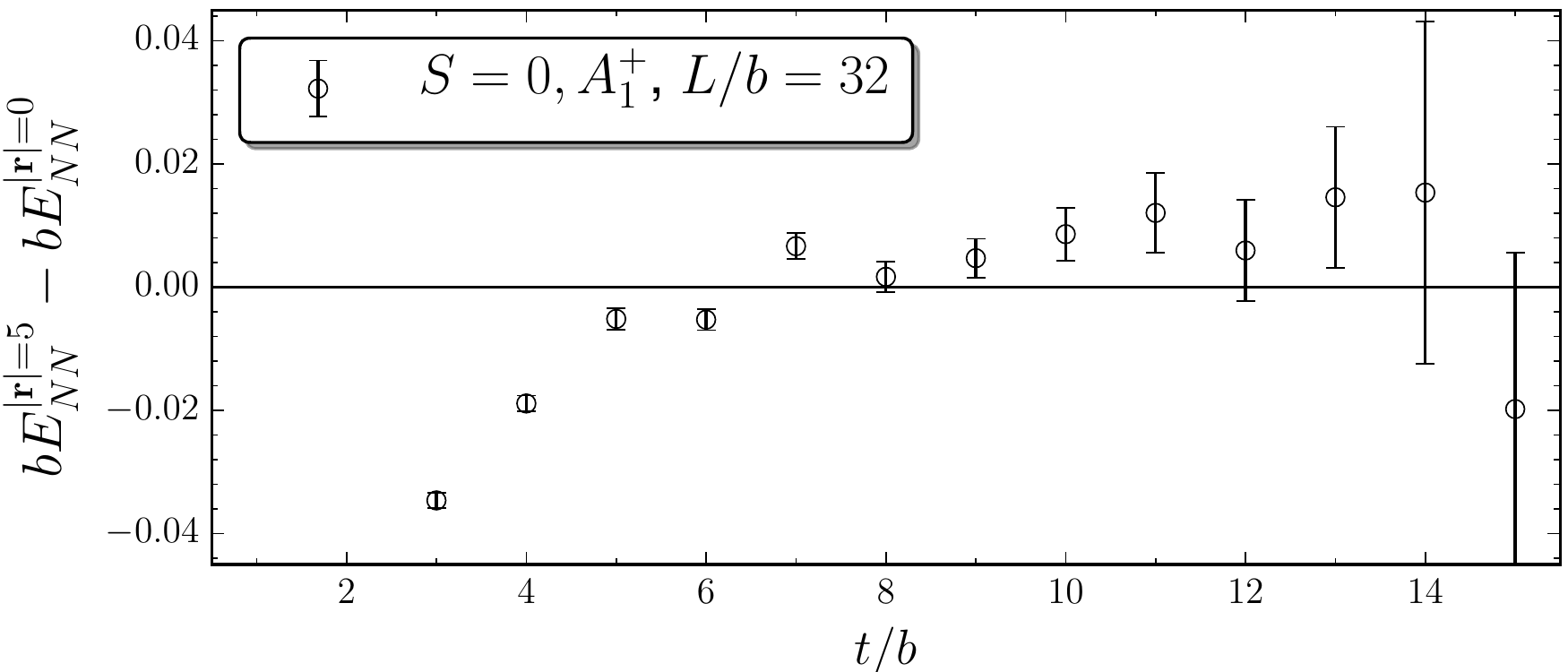}
\caption{\label{fig:neg_energy_states}Differences of effective masses using displaced vs. local sources for the lowest energy levels in the spin singlet, $A_1^+$ and spin triplet,$T_1^+$ channels on each volume. In all cases except for the spin singlet, $A_1^+$ channel with $L/b=24$, we find significantly distinct energy levels using the two different types of sources.}
\end{figure*}

We further elaborate on the plausibility of finding two negatively shifted energy states in both $S$-wave channels.
The large negatively shifted energy levels, determined with the local operators, are consistent with those in Ref.~\cite{Beane:2012vq}, which also used local operators.%
\footnote{Note, our results for the binding energies do differ somewhat from those in Ref.~\cite{Beane:2012vq} due in part to different strategies to determine the infinite volume binding energies from the finite volume energy levels.}  
The state closer to threshold (and additionally, the negative energy state near threshold in the $^1S_0$ channel) has strong overlap onto the non-local NN interpolating field, and has not been found in previous works. 
In Fig.~\ref{fig:neg_energy_states}, we plot the effective mass of the ratio correlation functions of the non-local two-nucleon interpolating fields divided by the local two-nucleon interpolating fields
\begin{align}
&R^{\Lambda}_{NN}(t) = C_{|\mathbf{r}| \neq 0}^{\Lambda}(t) / C_{|\mathbf{r}| = 0}^{\Lambda}(t)\, ,&
&C_{|\mathbf{r}|}^{\Lambda}(t) = \sum_n A_n^{\Lambda}({|\mathbf{r}|}) e^{-E_n^{\Lambda} t}\, .&
\end{align}
Both two-nucleon interpolating fields couple to the same tower of states, $E_n^{\Lambda} = 2m_N + \Delta_n^{\Lambda}$ and only differ in the relative size of their overlap factors, $A_n^{\Lambda}({|\mathbf{r}|})$.
In the spin triplet, $T_1^+$ channel, there is a clear plateau inconsistent with 0 for a relatively large time in fm (recall that each time step on these ensembles correspond to $a=0.1453(16)$~fm~\cite{Beane:2012vq}), indicating that the two correlation functions each give statistically distinct plateaus.
For the spin singlet, $A_1^+$ channel, a clear plateau is only observed on the larger volume.

In the long-time limit, the effective masses for all interpolating operators in this channel must asymptote to the ground state of the system. However, if the non-local interpolating field couples strongly to the state with small negative energy shift, and weakly to the state with large negative energy shift, and conversely for the local interpolating field, then at intermediate times the dominant contribution to each interpolating field will come from the two different states respectively.
This would manifest in the non-local interpolating field having an effective mass plateau at the smaller negative energy shift (the first excited state) at intermediate times.  
With enough statistics, one would eventually observe the effective mass ``collapse'' to the ground state and plateau again at the larger negative energy shift (the ground state).
Note that this is also true of all excited two-nucleon elastic scattering states. Because relative momentum is not a good quantum number, the non-zero relative momentum projections serve only to enhance the overlap of the interpolating operators with excited states relative to the ground state. We find this to work quite well due to the existence of distinct plateaus at intermediate times for the different relative momentum operators, however, in the very long time limit we expect all of these correlation functions to collapse onto a single ground state plateau.

This explanation for the two states with negative energies is consistent with the intuition that the wavefunction for a shallow bound state is more extended in space, and thus will have poor overlap with an interpolating field involving only two nucleons at the same point (a smeared out delta function in our case).
Conversely, the more deeply bound state would have a much more compact wave function and would thus overlap more poorly with a non-local interpolating field.
We finally emphasize that these findings are consistent with the two poles found using the ERE of our phase shift, even if we disregard the finite volume negative energy results from either the local or the non-local operators. We therefore find the most plausible explanation of these results to be the existence of two distinct bound states in the $T_1^+$ channel.

\section{Summary \label{sec:summary}}
This work presents the implementation of new two-nucleon interpolating fields which allow, for the first time, a robust determination of $\ell>0$ scattering phase shifts in the NN sector, which we have calculated in this exploratory work using unphysically heavy quark masses.
Further, this improved basis of interpolating operators are sensitive to additional states in the $S$-wave spectrum that were not found using only local operators and greater statistics.
This has been made possible by three previously unexploited tools. 
First was the development of displaced two-nucleon interpolating sources.
These are necessary to have appreciable overlap with partial waves beyond the S-wave as the $\ell\neq0$ orbital wavefunctions are zero at the origin.
Second was the use of momentum space sink operators that were not restricted to the simplest cubic irreps.
Finally, we applied the formalism for two-nucleon systems in a finite volume~\cite{Briceno:2013lba}, ignoring higher partial wave mixing, with notable success for the ${}^3P_2$ channel. 
This work represents the first crucial step towards the study of more challenging systems such as three-neutron interactions and the $S\rightarrow P$ wave parity violating $pp$ interaction.

\subsection*{Acknowledgments}
We would like to thank Ra\'{u}l Brice\~{n}o for extensive consultations regarding the finite volume formalism, as well as Dean Lee for discussions regarding the comparison between finite-volume bound state energies and unitarity. 
We would like to acknowledge W.~Detmold, R.~Edwards, D.~Richards and K.~Orginos for use of the JLab/W\&M configurations used in this work.
These calculations were performed with software built upon the Chroma software suite~\cite{Edwards:2004sx} and the optimized lattice QCD GPU library QUDA~\cite{Clark:2009wm,Babich:2011np}.
We also utilized the highly efficient HDF5 I/O Library~\cite{hdf5} with an interface to HDF5 in the USQCD Software Stack added with SciDAC 3 support~\cite{Kurth:2015mqa}.
We thank the Lawrence Livermore National Laboratory (LLNL) Multiprogrammatic and Institutional Computing program for the Grand Challenge allocation.  Our calculations were performed on the LLNL BG/Q supercomputer, Aztec cluster and Surface GPU cluster and on Edison at NERSC, the National Energy Research Scientific Computing Center (a DOE Office of Science User Facility supported by the Office of Science of the U.S. Department of Energy under Contract No. DE-AC02-05CH11231). We thank LLNL for funding from LDRD 13-ERD-023. This work was performed under the auspices of the U.S. Department of Energy by Lawrence Livermore National Laboratory under Contract DE-AC52-07NA27344.
This work was supported in part by the Office of Nuclear Physics in the US Department of Energy under grants KB0301052 (SciDAC), DE-SC00046548 (Berkeley), and DE-AC02-05CH11231.
The work of AWL was supported in part by the U.S. Department of Energy under contract DE-AC05-06OR23177, under which Jefferson Science Associates, LLC, manages and operates the Jefferson Lab, and by the U.S. DOE Early Career Award under contract DE-SC0012180.
\noindent

\bibliographystyle{apsrev} 
\bibliography{bibi} 

\begin{thebibliography}{80}
\expandafter\ifx\csname natexlab\endcsname\relax\def\natexlab#1{#1}\fi
\expandafter\ifx\csname bibnamefont\endcsname\relax
  \def\bibnamefont#1{#1}\fi
\expandafter\ifx\csname bibfnamefont\endcsname\relax
  \def\bibfnamefont#1{#1}\fi
\expandafter\ifx\csname citenamefont\endcsname\relax
  \def\citenamefont#1{#1}\fi
\expandafter\ifx\csname url\endcsname\relax
  \def\url#1{\texttt{#1}}\fi
\expandafter\ifx\csname urlprefix\endcsname\relax\def\urlprefix{URL }\fi
\providecommand{\bibinfo}[2]{#2}
\providecommand{\eprint}[2][]{\url{#2}}

\bibitem[{\citenamefont{Weinberg}(1990)}]{Weinberg:1990rz}
\bibinfo{author}{\bibfnamefont{S.}~\bibnamefont{Weinberg}},
  \bibinfo{journal}{Phys. Lett.} \textbf{\bibinfo{volume}{B251}},
  \bibinfo{pages}{288} (\bibinfo{year}{1990}).

\bibitem[{\citenamefont{Bedaque and van Kolck}(2002)}]{Bedaque:2002mn}
\bibinfo{author}{\bibfnamefont{P.~F.} \bibnamefont{Bedaque}} \bibnamefont{and}
  \bibinfo{author}{\bibfnamefont{U.}~\bibnamefont{van Kolck}},
  \bibinfo{journal}{Ann. Rev. Nucl. Part. Sci.} \textbf{\bibinfo{volume}{52}},
  \bibinfo{pages}{339} (\bibinfo{year}{2002}), \eprint{nucl-th/0203055}.

\bibitem[{\citenamefont{Epelbaum et~al.}(2009)\citenamefont{Epelbaum, Hammer,
  and Meissner}}]{Epelbaum:2008ga}
\bibinfo{author}{\bibfnamefont{E.}~\bibnamefont{Epelbaum}},
  \bibinfo{author}{\bibfnamefont{H.-W.} \bibnamefont{Hammer}},
  \bibnamefont{and} \bibinfo{author}{\bibfnamefont{U.-G.}
  \bibnamefont{Meissner}}, \bibinfo{journal}{Rev. Mod. Phys.}
  \textbf{\bibinfo{volume}{81}}, \bibinfo{pages}{1773} (\bibinfo{year}{2009}),
  \eprint{0811.1338}.

\bibitem[{\citenamefont{Haxton and Luu}(2002)}]{Haxton:2002kb}
\bibinfo{author}{\bibfnamefont{W.~C.} \bibnamefont{Haxton}} \bibnamefont{and}
  \bibinfo{author}{\bibfnamefont{T.}~\bibnamefont{Luu}},
  \bibinfo{journal}{Phys. Rev. Lett.} \textbf{\bibinfo{volume}{89}},
  \bibinfo{pages}{182503} (\bibinfo{year}{2002}), \eprint{nucl-th/0204072}.

\bibitem[{\citenamefont{Haxton}(2006)}]{Haxton:2006gw}
\bibinfo{author}{\bibfnamefont{W.~C.} \bibnamefont{Haxton}}
  (\bibinfo{year}{2006}), \eprint{nucl-th/0608017}.

\bibitem[{\citenamefont{Barrett et~al.}(2013)\citenamefont{Barrett, Navratil,
  and Vary}}]{Barrett:2013nh}
\bibinfo{author}{\bibfnamefont{B.~R.} \bibnamefont{Barrett}},
  \bibinfo{author}{\bibfnamefont{P.}~\bibnamefont{Navratil}}, \bibnamefont{and}
  \bibinfo{author}{\bibfnamefont{J.~P.} \bibnamefont{Vary}},
  \bibinfo{journal}{Prog.Part.Nucl.Phys.} \textbf{\bibinfo{volume}{69}},
  \bibinfo{pages}{131} (\bibinfo{year}{2013}).

\bibitem[{\citenamefont{Navratil et~al.}(2007)\citenamefont{Navratil,
  Gueorguiev, Vary, Ormand, and Nogga}}]{Navratil:2007we}
\bibinfo{author}{\bibfnamefont{P.}~\bibnamefont{Navratil}},
  \bibinfo{author}{\bibfnamefont{V.~G.} \bibnamefont{Gueorguiev}},
  \bibinfo{author}{\bibfnamefont{J.~P.} \bibnamefont{Vary}},
  \bibinfo{author}{\bibfnamefont{W.~E.} \bibnamefont{Ormand}},
  \bibnamefont{and} \bibinfo{author}{\bibfnamefont{A.}~\bibnamefont{Nogga}},
  \bibinfo{journal}{Phys. Rev. Lett.} \textbf{\bibinfo{volume}{99}},
  \bibinfo{pages}{042501} (\bibinfo{year}{2007}), \eprint{nucl-th/0701038}.

\bibitem[{\citenamefont{Gezerlis et~al.}(2013)\citenamefont{Gezerlis, Tews,
  Epelbaum, Gandolfi, Hebeler et~al.}}]{Gezerlis:2013ipa}
\bibinfo{author}{\bibfnamefont{A.}~\bibnamefont{Gezerlis}},
  \bibinfo{author}{\bibfnamefont{I.}~\bibnamefont{Tews}},
  \bibinfo{author}{\bibfnamefont{E.}~\bibnamefont{Epelbaum}},
  \bibinfo{author}{\bibfnamefont{S.}~\bibnamefont{Gandolfi}},
  \bibinfo{author}{\bibfnamefont{K.}~\bibnamefont{Hebeler}},
  \bibnamefont{et~al.}, \bibinfo{journal}{Phys.Rev.Lett.}
  \textbf{\bibinfo{volume}{111}}, \bibinfo{pages}{032501}
  (\bibinfo{year}{2013}), \eprint{1303.6243}.

\bibitem[{\citenamefont{Carlson et~al.}(2014)\citenamefont{Carlson, Gandolfi,
  Pederiva, Pieper, Schiavilla, Schmidt, and Wiringa}}]{Carlson:2014vla}
\bibinfo{author}{\bibfnamefont{J.}~\bibnamefont{Carlson}},
  \bibinfo{author}{\bibfnamefont{S.}~\bibnamefont{Gandolfi}},
  \bibinfo{author}{\bibfnamefont{F.}~\bibnamefont{Pederiva}},
  \bibinfo{author}{\bibfnamefont{S.~C.} \bibnamefont{Pieper}},
  \bibinfo{author}{\bibfnamefont{R.}~\bibnamefont{Schiavilla}},
  \bibinfo{author}{\bibfnamefont{K.~E.} \bibnamefont{Schmidt}},
  \bibnamefont{and} \bibinfo{author}{\bibfnamefont{R.~B.}
  \bibnamefont{Wiringa}} (\bibinfo{year}{2014}), \eprint{1412.3081}.

\bibitem[{\citenamefont{Pieper and Wiringa}(2001)}]{Pieper:2001mp}
\bibinfo{author}{\bibfnamefont{S.~C.} \bibnamefont{Pieper}} \bibnamefont{and}
  \bibinfo{author}{\bibfnamefont{R.~B.} \bibnamefont{Wiringa}},
  \bibinfo{journal}{Ann.Rev.Nucl.Part.Sci.} \textbf{\bibinfo{volume}{51}},
  \bibinfo{pages}{53} (\bibinfo{year}{2001}), \eprint{nucl-th/0103005}.

\bibitem[{\citenamefont{Hammer et~al.}(2013)\citenamefont{Hammer, Nogga, and
  Schwenk}}]{Hammer:2012id}
\bibinfo{author}{\bibfnamefont{H.-W.} \bibnamefont{Hammer}},
  \bibinfo{author}{\bibfnamefont{A.}~\bibnamefont{Nogga}}, \bibnamefont{and}
  \bibinfo{author}{\bibfnamefont{A.}~\bibnamefont{Schwenk}},
  \bibinfo{journal}{Rev.Mod.Phys.} \textbf{\bibinfo{volume}{85}},
  \bibinfo{pages}{197} (\bibinfo{year}{2013}), \eprint{1210.4273}.

\bibitem[{\citenamefont{Epelbaum et~al.}(2011)\citenamefont{Epelbaum, Krebs,
  Lee, and Meissner}}]{Epelbaum:2011md}
\bibinfo{author}{\bibfnamefont{E.}~\bibnamefont{Epelbaum}},
  \bibinfo{author}{\bibfnamefont{H.}~\bibnamefont{Krebs}},
  \bibinfo{author}{\bibfnamefont{D.}~\bibnamefont{Lee}}, \bibnamefont{and}
  \bibinfo{author}{\bibfnamefont{U.-G.} \bibnamefont{Meissner}},
  \bibinfo{journal}{Phys.Rev.Lett.} \textbf{\bibinfo{volume}{106}},
  \bibinfo{pages}{192501} (\bibinfo{year}{2011}), \eprint{1101.2547}.

\bibitem[{\citenamefont{Barnea et~al.}(2015)\citenamefont{Barnea, Contessi,
  Gazit, Pederiva, and van Kolck}}]{Barnea:2013uqa}
\bibinfo{author}{\bibfnamefont{N.}~\bibnamefont{Barnea}},
  \bibinfo{author}{\bibfnamefont{L.}~\bibnamefont{Contessi}},
  \bibinfo{author}{\bibfnamefont{D.}~\bibnamefont{Gazit}},
  \bibinfo{author}{\bibfnamefont{F.}~\bibnamefont{Pederiva}}, \bibnamefont{and}
  \bibinfo{author}{\bibfnamefont{U.}~\bibnamefont{van Kolck}},
  \bibinfo{journal}{Phys.Rev.Lett.} \textbf{\bibinfo{volume}{114}},
  \bibinfo{pages}{052501} (\bibinfo{year}{2015}), \eprint{1311.4966}.

\bibitem[{\citenamefont{Kirscher et~al.}(2015)\citenamefont{Kirscher, Barnea,
  Gazit, Pederiva, and van Kolck}}]{Kirscher:2015yda}
\bibinfo{author}{\bibfnamefont{J.}~\bibnamefont{Kirscher}},
  \bibinfo{author}{\bibfnamefont{N.}~\bibnamefont{Barnea}},
  \bibinfo{author}{\bibfnamefont{D.}~\bibnamefont{Gazit}},
  \bibinfo{author}{\bibfnamefont{F.}~\bibnamefont{Pederiva}}, \bibnamefont{and}
  \bibinfo{author}{\bibfnamefont{U.}~\bibnamefont{van Kolck}}
  (\bibinfo{year}{2015}), \eprint{1506.09048}.

\bibitem[{\citenamefont{Balantekin et~al.}(2014)}]{Balantekin:2014opa}
\bibinfo{author}{\bibfnamefont{A.~B.} \bibnamefont{Balantekin}}
  \bibnamefont{et~al.}, \bibinfo{journal}{Mod. Phys. Lett.}
  \textbf{\bibinfo{volume}{A29}}, \bibinfo{pages}{1430010}
  (\bibinfo{year}{2014}), \eprint{1401.6435}.

\bibitem[{\citenamefont{Nemura et~al.}(2009)\citenamefont{Nemura, Ishii, Aoki,
  and Hatsuda}}]{Nemura:2008sp}
\bibinfo{author}{\bibfnamefont{H.}~\bibnamefont{Nemura}},
  \bibinfo{author}{\bibfnamefont{N.}~\bibnamefont{Ishii}},
  \bibinfo{author}{\bibfnamefont{S.}~\bibnamefont{Aoki}}, \bibnamefont{and}
  \bibinfo{author}{\bibfnamefont{T.}~\bibnamefont{Hatsuda}},
  \bibinfo{journal}{Phys.Lett.} \textbf{\bibinfo{volume}{B673}},
  \bibinfo{pages}{136} (\bibinfo{year}{2009}), \eprint{0806.1094}.

\bibitem[{\citenamefont{Beane et~al.}(2011)}]{Beane:2010hg}
\bibinfo{author}{\bibfnamefont{S.}~\bibnamefont{Beane}} \bibnamefont{et~al.}
  (\bibinfo{collaboration}{NPLQCD}), \bibinfo{journal}{Phys.Rev.Lett.}
  \textbf{\bibinfo{volume}{106}}, \bibinfo{pages}{162001}
  (\bibinfo{year}{2011}), \eprint{1012.3812}.

\bibitem[{\citenamefont{Inoue et~al.}(2011)}]{Inoue:2010es}
\bibinfo{author}{\bibfnamefont{T.}~\bibnamefont{Inoue}} \bibnamefont{et~al.}
  (\bibinfo{collaboration}{HAL QCD}), \bibinfo{journal}{Phys.Rev.Lett.}
  \textbf{\bibinfo{volume}{106}}, \bibinfo{pages}{162002}
  (\bibinfo{year}{2011}), \eprint{1012.5928}.

\bibitem[{\citenamefont{Beane et~al.}(2012{\natexlab{a}})\citenamefont{Beane,
  Chang, Cohen, Detmold, Lin et~al.}}]{Beane:2012ey}
\bibinfo{author}{\bibfnamefont{S.}~\bibnamefont{Beane}},
  \bibinfo{author}{\bibfnamefont{E.}~\bibnamefont{Chang}},
  \bibinfo{author}{\bibfnamefont{S.}~\bibnamefont{Cohen}},
  \bibinfo{author}{\bibfnamefont{W.}~\bibnamefont{Detmold}},
  \bibinfo{author}{\bibfnamefont{H.-W.} \bibnamefont{Lin}},
  \bibnamefont{et~al.}, \bibinfo{journal}{Phys.Rev.Lett.}
  \textbf{\bibinfo{volume}{109}}, \bibinfo{pages}{172001}
  (\bibinfo{year}{2012}{\natexlab{a}}), \eprint{1204.3606}.

\bibitem[{\citenamefont{Adamczyk et~al.}(2015)}]{Adamczyk:2014vca}
\bibinfo{author}{\bibfnamefont{L.}~\bibnamefont{Adamczyk}} \bibnamefont{et~al.}
  (\bibinfo{collaboration}{STAR}), \bibinfo{journal}{Phys. Rev. Lett.}
  \textbf{\bibinfo{volume}{114}}, \bibinfo{pages}{022301}
  (\bibinfo{year}{2015}), \eprint{1408.4360}.

\bibitem[{\citenamefont{Adam et~al.}(2015)}]{Adam:2015nca}
\bibinfo{author}{\bibfnamefont{J.}~\bibnamefont{Adam}} \bibnamefont{et~al.}
  (\bibinfo{collaboration}{ALICE}) (\bibinfo{year}{2015}), \eprint{1506.07499}.

\bibitem[{\citenamefont{Wasem}(2012)}]{Wasem:2011zz}
\bibinfo{author}{\bibfnamefont{J.}~\bibnamefont{Wasem}},
  \bibinfo{journal}{Phys.Rev.} \textbf{\bibinfo{volume}{C85}},
  \bibinfo{pages}{022501} (\bibinfo{year}{2012}), \eprint{1108.1151}.

\bibitem[{\citenamefont{Beane et~al.}(2015)\citenamefont{Beane, Chang, Detmold,
  Orginos, Parreño et~al.}}]{Beane:2015yha}
\bibinfo{author}{\bibfnamefont{S.~R.} \bibnamefont{Beane}},
  \bibinfo{author}{\bibfnamefont{E.}~\bibnamefont{Chang}},
  \bibinfo{author}{\bibfnamefont{W.}~\bibnamefont{Detmold}},
  \bibinfo{author}{\bibfnamefont{K.}~\bibnamefont{Orginos}},
  \bibinfo{author}{\bibfnamefont{A.}~\bibnamefont{Parreño}},
  \bibnamefont{et~al.} (\bibinfo{year}{2015}), \eprint{1505.02422}.

\bibitem[{\citenamefont{Lellouch and Luscher}(2001)}]{Lellouch:2000pv}
\bibinfo{author}{\bibfnamefont{L.}~\bibnamefont{Lellouch}} \bibnamefont{and}
  \bibinfo{author}{\bibfnamefont{M.}~\bibnamefont{Luscher}},
  \bibinfo{journal}{Commun.Math.Phys.} \textbf{\bibinfo{volume}{219}},
  \bibinfo{pages}{31} (\bibinfo{year}{2001}), \eprint{hep-lat/0003023}.

\bibitem[{\citenamefont{Briceno
  et~al.}(2015{\natexlab{a}})\citenamefont{Briceno, Hansen, and
  Walker-Loud}}]{Briceno:2014uqa}
\bibinfo{author}{\bibfnamefont{R.}~\bibnamefont{Briceno}},
  \bibinfo{author}{\bibfnamefont{M.~T.} \bibnamefont{Hansen}},
  \bibnamefont{and}
  \bibinfo{author}{\bibfnamefont{A.}~\bibnamefont{Walker-Loud}},
  \bibinfo{journal}{Phys.Rev.} \textbf{\bibinfo{volume}{D91}},
  \bibinfo{pages}{034501} (\bibinfo{year}{2015}{\natexlab{a}}),
  \eprint{1406.5965}.

\bibitem[{\citenamefont{Briceno and Hansen}(2015)}]{Briceno:2015csa}
\bibinfo{author}{\bibfnamefont{R.}~\bibnamefont{Briceno}} \bibnamefont{and}
  \bibinfo{author}{\bibfnamefont{M.~T.} \bibnamefont{Hansen}}
  (\bibinfo{year}{2015}), \eprint{1502.04314}.

\bibitem[{\citenamefont{Briceno
  et~al.}(2015{\natexlab{b}})\citenamefont{Briceno, Dudek, Edwards, Shultz,
  Thomas, and Wilson}}]{Briceno:2015dca}
\bibinfo{author}{\bibfnamefont{R.~A.} \bibnamefont{Briceno}},
  \bibinfo{author}{\bibfnamefont{J.~J.} \bibnamefont{Dudek}},
  \bibinfo{author}{\bibfnamefont{R.~G.} \bibnamefont{Edwards}},
  \bibinfo{author}{\bibfnamefont{C.~J.} \bibnamefont{Shultz}},
  \bibinfo{author}{\bibfnamefont{C.~E.} \bibnamefont{Thomas}},
  \bibnamefont{and} \bibinfo{author}{\bibfnamefont{D.~J.} \bibnamefont{Wilson}}
  (\bibinfo{year}{2015}{\natexlab{b}}), \eprint{1507.06622}.

\bibitem[{\citenamefont{Dudek et~al.}(2012)\citenamefont{Dudek, Edwards, and
  Thomas}}]{Dudek:2012gj}
\bibinfo{author}{\bibfnamefont{J.~J.} \bibnamefont{Dudek}},
  \bibinfo{author}{\bibfnamefont{R.~G.} \bibnamefont{Edwards}},
  \bibnamefont{and} \bibinfo{author}{\bibfnamefont{C.~E.}
  \bibnamefont{Thomas}}, \bibinfo{journal}{Phys.Rev.}
  \textbf{\bibinfo{volume}{D86}}, \bibinfo{pages}{034031}
  (\bibinfo{year}{2012}), \eprint{1203.6041}.

\bibitem[{\citenamefont{Dudek et~al.}(2013)\citenamefont{Dudek, Edwards, and
  Thomas}}]{Dudek:2012xn}
\bibinfo{author}{\bibfnamefont{J.~J.} \bibnamefont{Dudek}},
  \bibinfo{author}{\bibfnamefont{R.~G.} \bibnamefont{Edwards}},
  \bibnamefont{and} \bibinfo{author}{\bibfnamefont{C.~E.} \bibnamefont{Thomas}}
  (\bibinfo{collaboration}{Hadron Spectrum}), \bibinfo{journal}{Phys.Rev.}
  \textbf{\bibinfo{volume}{D87}}, \bibinfo{pages}{034505}
  (\bibinfo{year}{2013}), \eprint{1212.0830}.

\bibitem[{\citenamefont{Pelissier and Alexandru}(2013)}]{Pelissier:2012pi}
\bibinfo{author}{\bibfnamefont{C.}~\bibnamefont{Pelissier}} \bibnamefont{and}
  \bibinfo{author}{\bibfnamefont{A.}~\bibnamefont{Alexandru}},
  \bibinfo{journal}{Phys. Rev.} \textbf{\bibinfo{volume}{D87}},
  \bibinfo{pages}{014503} (\bibinfo{year}{2013}), \eprint{1211.0092}.

\bibitem[{\citenamefont{Lang et~al.}(2014)\citenamefont{Lang, Leskovec, Mohler,
  and Prelovsek}}]{Lang:2014tia}
\bibinfo{author}{\bibfnamefont{C.~B.} \bibnamefont{Lang}},
  \bibinfo{author}{\bibfnamefont{L.}~\bibnamefont{Leskovec}},
  \bibinfo{author}{\bibfnamefont{D.}~\bibnamefont{Mohler}}, \bibnamefont{and}
  \bibinfo{author}{\bibfnamefont{S.}~\bibnamefont{Prelovsek}},
  \bibinfo{journal}{JHEP} \textbf{\bibinfo{volume}{04}}, \bibinfo{pages}{162}
  (\bibinfo{year}{2014}), \eprint{1401.2088}.

\bibitem[{\citenamefont{Bali et~al.}(2016)\citenamefont{Bali, Collins, Cox,
  Donald, Göckeler, Lang, and Schäfer}}]{Bali:2015gji}
\bibinfo{author}{\bibfnamefont{G.~S.} \bibnamefont{Bali}},
  \bibinfo{author}{\bibfnamefont{S.}~\bibnamefont{Collins}},
  \bibinfo{author}{\bibfnamefont{A.}~\bibnamefont{Cox}},
  \bibinfo{author}{\bibfnamefont{G.}~\bibnamefont{Donald}},
  \bibinfo{author}{\bibfnamefont{M.}~\bibnamefont{Göckeler}},
  \bibinfo{author}{\bibfnamefont{C.~B.} \bibnamefont{Lang}}, \bibnamefont{and}
  \bibinfo{author}{\bibfnamefont{A.}~\bibnamefont{Schäfer}}
  (\bibinfo{collaboration}{RQCD}), \bibinfo{journal}{Phys. Rev.}
  \textbf{\bibinfo{volume}{D93}}, \bibinfo{pages}{054509}
  (\bibinfo{year}{2016}), \eprint{1512.08678}.

\bibitem[{\citenamefont{Bulava et~al.}(2016)\citenamefont{Bulava, Fahy, Hörz,
  Juge, Morningstar, and Wong}}]{Bulava:2016mks}
\bibinfo{author}{\bibfnamefont{J.}~\bibnamefont{Bulava}},
  \bibinfo{author}{\bibfnamefont{B.}~\bibnamefont{Fahy}},
  \bibinfo{author}{\bibfnamefont{B.}~\bibnamefont{Hörz}},
  \bibinfo{author}{\bibfnamefont{K.~J.} \bibnamefont{Juge}},
  \bibinfo{author}{\bibfnamefont{C.}~\bibnamefont{Morningstar}},
  \bibnamefont{and} \bibinfo{author}{\bibfnamefont{C.~H.} \bibnamefont{Wong}},
  \bibinfo{journal}{Nucl. Phys.} \textbf{\bibinfo{volume}{B910}},
  \bibinfo{pages}{842} (\bibinfo{year}{2016}), \eprint{1604.05593}.

\bibitem[{\citenamefont{Guo et~al.}(2016)\citenamefont{Guo, Alexandru, Molina,
  and Döring}}]{Guo:2016zos}
\bibinfo{author}{\bibfnamefont{D.}~\bibnamefont{Guo}},
  \bibinfo{author}{\bibfnamefont{A.}~\bibnamefont{Alexandru}},
  \bibinfo{author}{\bibfnamefont{R.}~\bibnamefont{Molina}}, \bibnamefont{and}
  \bibinfo{author}{\bibfnamefont{M.}~\bibnamefont{Döring}},
  \bibinfo{journal}{Phys. Rev.} \textbf{\bibinfo{volume}{D94}},
  \bibinfo{pages}{034501} (\bibinfo{year}{2016}), \eprint{1605.03993}.

\bibitem[{\citenamefont{Feng et~al.}(2011)\citenamefont{Feng, Jansen, and
  Renner}}]{Feng:2010es}
\bibinfo{author}{\bibfnamefont{X.}~\bibnamefont{Feng}},
  \bibinfo{author}{\bibfnamefont{K.}~\bibnamefont{Jansen}}, \bibnamefont{and}
  \bibinfo{author}{\bibfnamefont{D.~B.} \bibnamefont{Renner}},
  \bibinfo{journal}{Phys.Rev.} \textbf{\bibinfo{volume}{D83}},
  \bibinfo{pages}{094505} (\bibinfo{year}{2011}), \eprint{1011.5288}.

\bibitem[{\citenamefont{Beane et~al.}(2012{\natexlab{b}})\citenamefont{Beane,
  Chang, Detmold, Lin, Luu, Orginos, Parreno, Savage, Torok, and
  Walker-Loud}}]{Beane:2011sc}
\bibinfo{author}{\bibfnamefont{S.~R.} \bibnamefont{Beane}},
  \bibinfo{author}{\bibfnamefont{E.}~\bibnamefont{Chang}},
  \bibinfo{author}{\bibfnamefont{W.}~\bibnamefont{Detmold}},
  \bibinfo{author}{\bibfnamefont{H.~W.} \bibnamefont{Lin}},
  \bibinfo{author}{\bibfnamefont{T.~C.} \bibnamefont{Luu}},
  \bibinfo{author}{\bibfnamefont{K.}~\bibnamefont{Orginos}},
  \bibinfo{author}{\bibfnamefont{A.}~\bibnamefont{Parreno}},
  \bibinfo{author}{\bibfnamefont{M.~J.} \bibnamefont{Savage}},
  \bibinfo{author}{\bibfnamefont{A.}~\bibnamefont{Torok}}, \bibnamefont{and}
  \bibinfo{author}{\bibfnamefont{A.}~\bibnamefont{Walker-Loud}}
  (\bibinfo{collaboration}{NPLQCD}), \bibinfo{journal}{Phys. Rev.}
  \textbf{\bibinfo{volume}{D85}}, \bibinfo{pages}{034505}
  (\bibinfo{year}{2012}{\natexlab{b}}), \eprint{1107.5023}.

\bibitem[{\citenamefont{Luscher}(1986)}]{Luscher:1986pf}
\bibinfo{author}{\bibfnamefont{M.}~\bibnamefont{Luscher}},
  \bibinfo{journal}{Commun.Math.Phys.} \textbf{\bibinfo{volume}{105}},
  \bibinfo{pages}{153} (\bibinfo{year}{1986}).

\bibitem[{\citenamefont{Luscher}(1991)}]{Luscher:1990ux}
\bibinfo{author}{\bibfnamefont{M.}~\bibnamefont{Luscher}},
  \bibinfo{journal}{Nucl.Phys.} \textbf{\bibinfo{volume}{B354}},
  \bibinfo{pages}{531} (\bibinfo{year}{1991}).

\bibitem[{\citenamefont{Rummukainen and Gottlieb}(1995)}]{Rummukainen:1995vs}
\bibinfo{author}{\bibfnamefont{K.}~\bibnamefont{Rummukainen}} \bibnamefont{and}
  \bibinfo{author}{\bibfnamefont{S.~A.} \bibnamefont{Gottlieb}},
  \bibinfo{journal}{Nucl. Phys.} \textbf{\bibinfo{volume}{B450}},
  \bibinfo{pages}{397} (\bibinfo{year}{1995}), \eprint{hep-lat/9503028}.

\bibitem[{\citenamefont{Kim et~al.}(2005)\citenamefont{Kim, Sachrajda, and
  Sharpe}}]{Kim:2005gf}
\bibinfo{author}{\bibfnamefont{C.}~\bibnamefont{Kim}},
  \bibinfo{author}{\bibfnamefont{C.}~\bibnamefont{Sachrajda}},
  \bibnamefont{and} \bibinfo{author}{\bibfnamefont{S.~R.}
  \bibnamefont{Sharpe}}, \bibinfo{journal}{Nucl.Phys.}
  \textbf{\bibinfo{volume}{B727}}, \bibinfo{pages}{218} (\bibinfo{year}{2005}),
  \eprint{hep-lat/0507006}.

\bibitem[{\citenamefont{Christ et~al.}(2005)\citenamefont{Christ, Kim, and
  Yamazaki}}]{Christ:2005gi}
\bibinfo{author}{\bibfnamefont{N.~H.} \bibnamefont{Christ}},
  \bibinfo{author}{\bibfnamefont{C.}~\bibnamefont{Kim}}, \bibnamefont{and}
  \bibinfo{author}{\bibfnamefont{T.}~\bibnamefont{Yamazaki}},
  \bibinfo{journal}{Phys.Rev.} \textbf{\bibinfo{volume}{D72}},
  \bibinfo{pages}{114506} (\bibinfo{year}{2005}), \eprint{hep-lat/0507009}.

\bibitem[{\citenamefont{He et~al.}(2005)\citenamefont{He, Feng, and
  Liu}}]{He:2005ey}
\bibinfo{author}{\bibfnamefont{S.}~\bibnamefont{He}},
  \bibinfo{author}{\bibfnamefont{X.}~\bibnamefont{Feng}}, \bibnamefont{and}
  \bibinfo{author}{\bibfnamefont{C.}~\bibnamefont{Liu}},
  \bibinfo{journal}{JHEP} \textbf{\bibinfo{volume}{0507}}, \bibinfo{pages}{011}
  (\bibinfo{year}{2005}), \eprint{hep-lat/0504019}.

\bibitem[{\citenamefont{Briceno and
  Davoudi}(2013{\natexlab{a}})}]{Briceno:2012yi}
\bibinfo{author}{\bibfnamefont{R.~A.} \bibnamefont{Briceno}} \bibnamefont{and}
  \bibinfo{author}{\bibfnamefont{Z.}~\bibnamefont{Davoudi}},
  \bibinfo{journal}{Phys. Rev.} \textbf{\bibinfo{volume}{D88}},
  \bibinfo{pages}{094507} (\bibinfo{year}{2013}{\natexlab{a}}),
  \eprint{1204.1110}.

\bibitem[{\citenamefont{Hansen and Sharpe}(2012)}]{Hansen:2012tf}
\bibinfo{author}{\bibfnamefont{M.~T.} \bibnamefont{Hansen}} \bibnamefont{and}
  \bibinfo{author}{\bibfnamefont{S.~R.} \bibnamefont{Sharpe}},
  \bibinfo{journal}{Phys.Rev.} \textbf{\bibinfo{volume}{D86}},
  \bibinfo{pages}{016007} (\bibinfo{year}{2012}), \eprint{1204.0826}.

\bibitem[{\citenamefont{Briceno}(2014)}]{Briceno:2014oea}
\bibinfo{author}{\bibfnamefont{R.~A.} \bibnamefont{Briceno}},
  \bibinfo{journal}{Phys.Rev.} \textbf{\bibinfo{volume}{D89}},
  \bibinfo{pages}{074507} (\bibinfo{year}{2014}), \eprint{1401.3312}.

\bibitem[{\citenamefont{Dudek et~al.}(2014)\citenamefont{Dudek, Edwards,
  Thomas, and Wilson}}]{Dudek:2014qha}
\bibinfo{author}{\bibfnamefont{J.~J.} \bibnamefont{Dudek}},
  \bibinfo{author}{\bibfnamefont{R.~G.} \bibnamefont{Edwards}},
  \bibinfo{author}{\bibfnamefont{C.~E.} \bibnamefont{Thomas}},
  \bibnamefont{and} \bibinfo{author}{\bibfnamefont{D.~J.} \bibnamefont{Wilson}}
  (\bibinfo{collaboration}{Hadron Spectrum}), \bibinfo{journal}{Phys.Rev.Lett.}
  \textbf{\bibinfo{volume}{113}}, \bibinfo{pages}{182001}
  (\bibinfo{year}{2014}), \eprint{1406.4158}.

\bibitem[{\citenamefont{Wilson et~al.}(2014)\citenamefont{Wilson, Dudek,
  Edwards, and Thomas}}]{Wilson:2014cna}
\bibinfo{author}{\bibfnamefont{D.~J.} \bibnamefont{Wilson}},
  \bibinfo{author}{\bibfnamefont{J.~J.} \bibnamefont{Dudek}},
  \bibinfo{author}{\bibfnamefont{R.~G.} \bibnamefont{Edwards}},
  \bibnamefont{and} \bibinfo{author}{\bibfnamefont{C.~E.} \bibnamefont{Thomas}}
  (\bibinfo{year}{2014}), \eprint{1411.2004}.

\bibitem[{\citenamefont{Wilson et~al.}(2015)\citenamefont{Wilson, Briceno,
  Dudek, Edwards, and Thomas}}]{Wilson:2015dqa}
\bibinfo{author}{\bibfnamefont{D.~J.} \bibnamefont{Wilson}},
  \bibinfo{author}{\bibfnamefont{R.~A.} \bibnamefont{Briceno}},
  \bibinfo{author}{\bibfnamefont{J.~J.} \bibnamefont{Dudek}},
  \bibinfo{author}{\bibfnamefont{R.~G.} \bibnamefont{Edwards}},
  \bibnamefont{and} \bibinfo{author}{\bibfnamefont{C.~E.} \bibnamefont{Thomas}}
  (\bibinfo{year}{2015}), \eprint{1507.02599}.

\bibitem[{\citenamefont{Walker-Loud}(2014)}]{Walker-Loud:2014iea}
\bibinfo{author}{\bibfnamefont{A.}~\bibnamefont{Walker-Loud}},
  \bibinfo{journal}{PoS} \textbf{\bibinfo{volume}{LATTICE2013}},
  \bibinfo{pages}{013} (\bibinfo{year}{2014}), \eprint{1401.8259}.

\bibitem[{\citenamefont{Briceno
  et~al.}(2015{\natexlab{c}})\citenamefont{Briceno, Davoudi, and
  Luu}}]{Briceno:2014tqa}
\bibinfo{author}{\bibfnamefont{R.~A.} \bibnamefont{Briceno}},
  \bibinfo{author}{\bibfnamefont{Z.}~\bibnamefont{Davoudi}}, \bibnamefont{and}
  \bibinfo{author}{\bibfnamefont{T.~C.} \bibnamefont{Luu}},
  \bibinfo{journal}{J.Phys.} \textbf{\bibinfo{volume}{G42}},
  \bibinfo{pages}{023101} (\bibinfo{year}{2015}{\natexlab{c}}),
  \eprint{1406.5673}.

\bibitem[{\citenamefont{Fukugita et~al.}(1995)\citenamefont{Fukugita,
  Kuramashi, Okawa, Mino, and Ukawa}}]{Fukugita:1994ve}
\bibinfo{author}{\bibfnamefont{M.}~\bibnamefont{Fukugita}},
  \bibinfo{author}{\bibfnamefont{Y.}~\bibnamefont{Kuramashi}},
  \bibinfo{author}{\bibfnamefont{M.}~\bibnamefont{Okawa}},
  \bibinfo{author}{\bibfnamefont{H.}~\bibnamefont{Mino}}, \bibnamefont{and}
  \bibinfo{author}{\bibfnamefont{A.}~\bibnamefont{Ukawa}},
  \bibinfo{journal}{Phys.Rev.} \textbf{\bibinfo{volume}{D52}},
  \bibinfo{pages}{3003} (\bibinfo{year}{1995}), \eprint{hep-lat/9501024}.

\bibitem[{\citenamefont{Beane et~al.}(2006)\citenamefont{Beane, Bedaque,
  Orginos, and Savage}}]{Beane:2006mx}
\bibinfo{author}{\bibfnamefont{S.}~\bibnamefont{Beane}},
  \bibinfo{author}{\bibfnamefont{P.}~\bibnamefont{Bedaque}},
  \bibinfo{author}{\bibfnamefont{K.}~\bibnamefont{Orginos}}, \bibnamefont{and}
  \bibinfo{author}{\bibfnamefont{M.}~\bibnamefont{Savage}},
  \bibinfo{journal}{Phys.Rev.Lett.} \textbf{\bibinfo{volume}{97}},
  \bibinfo{pages}{012001} (\bibinfo{year}{2006}), \eprint{hep-lat/0602010}.

\bibitem[{\citenamefont{Beane et~al.}(2013{\natexlab{a}})}]{Beane:2013br}
\bibinfo{author}{\bibfnamefont{S.}~\bibnamefont{Beane}} \bibnamefont{et~al.}
  (\bibinfo{collaboration}{NPLQCD Collaboration}), \bibinfo{journal}{Phys.Rev.}
  \textbf{\bibinfo{volume}{C88}}, \bibinfo{pages}{024003}
  (\bibinfo{year}{2013}{\natexlab{a}}), \eprint{1301.5790}.

\bibitem[{\citenamefont{Beane et~al.}(2012{\natexlab{c}})}]{Beane:2011iw}
\bibinfo{author}{\bibfnamefont{S.}~\bibnamefont{Beane}} \bibnamefont{et~al.}
  (\bibinfo{collaboration}{NPLQCD Collaboration}), \bibinfo{journal}{Phys.Rev.}
  \textbf{\bibinfo{volume}{D85}}, \bibinfo{pages}{054511}
  (\bibinfo{year}{2012}{\natexlab{c}}), \eprint{1109.2889}.

\bibitem[{\citenamefont{Aoki et~al.}(2012)\citenamefont{Aoki, Doi, Hatsuda,
  Ikeda, Inoue, Ishii, Murano, Nemura, and Sasaki}}]{Aoki:2012tk}
\bibinfo{author}{\bibfnamefont{S.}~\bibnamefont{Aoki}},
  \bibinfo{author}{\bibfnamefont{T.}~\bibnamefont{Doi}},
  \bibinfo{author}{\bibfnamefont{T.}~\bibnamefont{Hatsuda}},
  \bibinfo{author}{\bibfnamefont{Y.}~\bibnamefont{Ikeda}},
  \bibinfo{author}{\bibfnamefont{T.}~\bibnamefont{Inoue}},
  \bibinfo{author}{\bibfnamefont{N.}~\bibnamefont{Ishii}},
  \bibinfo{author}{\bibfnamefont{K.}~\bibnamefont{Murano}},
  \bibinfo{author}{\bibfnamefont{H.}~\bibnamefont{Nemura}}, \bibnamefont{and}
  \bibinfo{author}{\bibfnamefont{K.}~\bibnamefont{Sasaki}}
  (\bibinfo{collaboration}{HAL QCD}), \bibinfo{journal}{PTEP}
  \textbf{\bibinfo{volume}{2012}}, \bibinfo{pages}{01A105}
  (\bibinfo{year}{2012}), \eprint{1206.5088}.

\bibitem[{\citenamefont{Yamazaki et~al.}(2011)\citenamefont{Yamazaki,
  Kuramashi, and Ukawa}}]{Yamazaki:2011nd}
\bibinfo{author}{\bibfnamefont{T.}~\bibnamefont{Yamazaki}},
  \bibinfo{author}{\bibfnamefont{Y.}~\bibnamefont{Kuramashi}},
  \bibnamefont{and} \bibinfo{author}{\bibfnamefont{A.}~\bibnamefont{Ukawa}}
  (\bibinfo{collaboration}{PACS-CS Collaboration}),
  \bibinfo{journal}{Phys.Rev.} \textbf{\bibinfo{volume}{D84}},
  \bibinfo{pages}{054506} (\bibinfo{year}{2011}), \eprint{1105.1418}.

\bibitem[{\citenamefont{Beane et~al.}(2013{\natexlab{b}})}]{Beane:2012vq}
\bibinfo{author}{\bibfnamefont{S.}~\bibnamefont{Beane}} \bibnamefont{et~al.}
  (\bibinfo{collaboration}{NPLQCD}), \bibinfo{journal}{Phys.Rev.}
  \textbf{\bibinfo{volume}{D87}}, \bibinfo{pages}{034506}
  (\bibinfo{year}{2013}{\natexlab{b}}), \eprint{1206.5219}.

\bibitem[{\citenamefont{Yamazaki et~al.}(2012)\citenamefont{Yamazaki, Ishikawa,
  Kuramashi, and Ukawa}}]{Yamazaki:2012hi}
\bibinfo{author}{\bibfnamefont{T.}~\bibnamefont{Yamazaki}},
  \bibinfo{author}{\bibfnamefont{K.-i.} \bibnamefont{Ishikawa}},
  \bibinfo{author}{\bibfnamefont{Y.}~\bibnamefont{Kuramashi}},
  \bibnamefont{and} \bibinfo{author}{\bibfnamefont{A.}~\bibnamefont{Ukawa}},
  \bibinfo{journal}{Phys.Rev.} \textbf{\bibinfo{volume}{D86}},
  \bibinfo{pages}{074514} (\bibinfo{year}{2012}), \eprint{1207.4277}.

\bibitem[{\citenamefont{Yamazaki et~al.}(2015)\citenamefont{Yamazaki, Ishikawa,
  Kuramashi, and Ukawa}}]{Yamazaki:2015asa}
\bibinfo{author}{\bibfnamefont{T.}~\bibnamefont{Yamazaki}},
  \bibinfo{author}{\bibfnamefont{K.-i.} \bibnamefont{Ishikawa}},
  \bibinfo{author}{\bibfnamefont{Y.}~\bibnamefont{Kuramashi}},
  \bibnamefont{and} \bibinfo{author}{\bibfnamefont{A.}~\bibnamefont{Ukawa}}
  (\bibinfo{year}{2015}), \eprint{1502.04182}.

\bibitem[{\citenamefont{Murano et~al.}(2014)}]{Murano:2013xxa}
\bibinfo{author}{\bibfnamefont{K.}~\bibnamefont{Murano}} \bibnamefont{et~al.}
  (\bibinfo{collaboration}{HAL QCD}), \bibinfo{journal}{Phys.Lett.}
  \textbf{\bibinfo{volume}{B735}}, \bibinfo{pages}{19} (\bibinfo{year}{2014}),
  \eprint{1305.2293}.

\bibitem[{\citenamefont{Ishii et~al.}(2007)\citenamefont{Ishii, Aoki, and
  Hatsuda}}]{Ishii:2006ec}
\bibinfo{author}{\bibfnamefont{N.}~\bibnamefont{Ishii}},
  \bibinfo{author}{\bibfnamefont{S.}~\bibnamefont{Aoki}}, \bibnamefont{and}
  \bibinfo{author}{\bibfnamefont{T.}~\bibnamefont{Hatsuda}},
  \bibinfo{journal}{Phys.Rev.Lett.} \textbf{\bibinfo{volume}{99}},
  \bibinfo{pages}{022001} (\bibinfo{year}{2007}), \eprint{nucl-th/0611096}.

\bibitem[{\citenamefont{Kurth et~al.}(2013)\citenamefont{Kurth, Ishii, Doi,
  Aoki, and Hatsuda}}]{Kurth:2013tua}
\bibinfo{author}{\bibfnamefont{T.}~\bibnamefont{Kurth}},
  \bibinfo{author}{\bibfnamefont{N.}~\bibnamefont{Ishii}},
  \bibinfo{author}{\bibfnamefont{T.}~\bibnamefont{Doi}},
  \bibinfo{author}{\bibfnamefont{S.}~\bibnamefont{Aoki}}, \bibnamefont{and}
  \bibinfo{author}{\bibfnamefont{T.}~\bibnamefont{Hatsuda}},
  \bibinfo{journal}{JHEP} \textbf{\bibinfo{volume}{1312}}, \bibinfo{pages}{015}
  (\bibinfo{year}{2013}), \eprint{1305.4462}.

\bibitem[{\citenamefont{Hansen and Sharpe}(2015)}]{Hansen:2015zga}
\bibinfo{author}{\bibfnamefont{M.~T.} \bibnamefont{Hansen}} \bibnamefont{and}
  \bibinfo{author}{\bibfnamefont{S.~R.} \bibnamefont{Sharpe}}
  (\bibinfo{year}{2015}), \eprint{1504.04248}.

\bibitem[{\citenamefont{Meißner et~al.}(2015)\citenamefont{Meißner, Ríos, and
  Rusetsky}}]{Meissner:2014dea}
\bibinfo{author}{\bibfnamefont{U.-G.} \bibnamefont{Meißner}},
  \bibinfo{author}{\bibfnamefont{G.}~\bibnamefont{Ríos}}, \bibnamefont{and}
  \bibinfo{author}{\bibfnamefont{A.}~\bibnamefont{Rusetsky}},
  \bibinfo{journal}{Phys.Rev.Lett.} \textbf{\bibinfo{volume}{114}},
  \bibinfo{pages}{091602} (\bibinfo{year}{2015}), \eprint{1412.4969}.

\bibitem[{\citenamefont{Hansen and Sharpe}(2014)}]{Hansen:2014eka}
\bibinfo{author}{\bibfnamefont{M.~T.} \bibnamefont{Hansen}} \bibnamefont{and}
  \bibinfo{author}{\bibfnamefont{S.~R.} \bibnamefont{Sharpe}},
  \bibinfo{journal}{Phys.Rev.} \textbf{\bibinfo{volume}{D90}},
  \bibinfo{pages}{116003} (\bibinfo{year}{2014}), \eprint{1408.5933}.

\bibitem[{\citenamefont{Briceno and
  Davoudi}(2013{\natexlab{b}})}]{Briceno:2012rv}
\bibinfo{author}{\bibfnamefont{R.~A.} \bibnamefont{Briceno}} \bibnamefont{and}
  \bibinfo{author}{\bibfnamefont{Z.}~\bibnamefont{Davoudi}},
  \bibinfo{journal}{Phys.Rev.} \textbf{\bibinfo{volume}{D87}},
  \bibinfo{pages}{094507} (\bibinfo{year}{2013}{\natexlab{b}}),
  \eprint{1212.3398}.

\bibitem[{\citenamefont{Polejaeva and Rusetsky}(2012)}]{Polejaeva:2012ut}
\bibinfo{author}{\bibfnamefont{K.}~\bibnamefont{Polejaeva}} \bibnamefont{and}
  \bibinfo{author}{\bibfnamefont{A.}~\bibnamefont{Rusetsky}},
  \bibinfo{journal}{Eur.Phys.J.} \textbf{\bibinfo{volume}{A48}},
  \bibinfo{pages}{67} (\bibinfo{year}{2012}), \eprint{1203.1241}.

\bibitem[{\citenamefont{Doi et~al.}(2012)\citenamefont{Doi, Aoki, Hatsuda,
  Ikeda, Inoue, Ishii, Murano, Nemura, and Sasaki}}]{Doi:2011gq}
\bibinfo{author}{\bibfnamefont{T.}~\bibnamefont{Doi}},
  \bibinfo{author}{\bibfnamefont{S.}~\bibnamefont{Aoki}},
  \bibinfo{author}{\bibfnamefont{T.}~\bibnamefont{Hatsuda}},
  \bibinfo{author}{\bibfnamefont{Y.}~\bibnamefont{Ikeda}},
  \bibinfo{author}{\bibfnamefont{T.}~\bibnamefont{Inoue}},
  \bibinfo{author}{\bibfnamefont{N.}~\bibnamefont{Ishii}},
  \bibinfo{author}{\bibfnamefont{K.}~\bibnamefont{Murano}},
  \bibinfo{author}{\bibfnamefont{H.}~\bibnamefont{Nemura}}, \bibnamefont{and}
  \bibinfo{author}{\bibfnamefont{K.}~\bibnamefont{Sasaki}}
  (\bibinfo{collaboration}{HAL QCD}), \bibinfo{journal}{Prog. Theor. Phys.}
  \textbf{\bibinfo{volume}{127}}, \bibinfo{pages}{723} (\bibinfo{year}{2012}),
  \eprint{1106.2276}.

\bibitem[{\citenamefont{Briceno et~al.}(2013)\citenamefont{Briceno, Davoudi,
  and Luu}}]{Briceno:2013lba}
\bibinfo{author}{\bibfnamefont{R.~A.} \bibnamefont{Briceno}},
  \bibinfo{author}{\bibfnamefont{Z.}~\bibnamefont{Davoudi}}, \bibnamefont{and}
  \bibinfo{author}{\bibfnamefont{T.~C.} \bibnamefont{Luu}},
  \bibinfo{journal}{Phys.Rev.} \textbf{\bibinfo{volume}{D88}},
  \bibinfo{pages}{034502} (\bibinfo{year}{2013}), \eprint{1305.4903}.

\bibitem[{\citenamefont{Basak et~al.}(2005)}]{Basak:2005ir}
\bibinfo{author}{\bibfnamefont{S.}~\bibnamefont{Basak}} \bibnamefont{et~al.}
  (\bibinfo{collaboration}{Lattice Hadron Physics (LHPC)}),
  \bibinfo{journal}{Phys.Rev.} \textbf{\bibinfo{volume}{D72}},
  \bibinfo{pages}{074501} (\bibinfo{year}{2005}), \eprint{hep-lat/0508018}.

\bibitem[{\citenamefont{Basak et~al.}(2007)\citenamefont{Basak, Edwards,
  Fleming, Juge, Lichtl et~al.}}]{Basak:2007kj}
\bibinfo{author}{\bibfnamefont{S.}~\bibnamefont{Basak}},
  \bibinfo{author}{\bibfnamefont{R.}~\bibnamefont{Edwards}},
  \bibinfo{author}{\bibfnamefont{G.}~\bibnamefont{Fleming}},
  \bibinfo{author}{\bibfnamefont{K.}~\bibnamefont{Juge}},
  \bibinfo{author}{\bibfnamefont{A.}~\bibnamefont{Lichtl}},
  \bibnamefont{et~al.}, \bibinfo{journal}{Phys.Rev.}
  \textbf{\bibinfo{volume}{D76}}, \bibinfo{pages}{074504}
  (\bibinfo{year}{2007}), \eprint{0709.0008}.

\bibitem[{\citenamefont{Morningstar et~al.}(2013)\citenamefont{Morningstar,
  Bulava, Fahy, Foley, Jhang et~al.}}]{Morningstar:2013bda}
\bibinfo{author}{\bibfnamefont{C.}~\bibnamefont{Morningstar}},
  \bibinfo{author}{\bibfnamefont{J.}~\bibnamefont{Bulava}},
  \bibinfo{author}{\bibfnamefont{B.}~\bibnamefont{Fahy}},
  \bibinfo{author}{\bibfnamefont{J.}~\bibnamefont{Foley}},
  \bibinfo{author}{\bibfnamefont{Y.}~\bibnamefont{Jhang}},
  \bibnamefont{et~al.}, \bibinfo{journal}{Phys.Rev.}
  \textbf{\bibinfo{volume}{D88}}, \bibinfo{pages}{014511}
  (\bibinfo{year}{2013}), \eprint{1303.6816}.

\bibitem[{\citenamefont{Dudek et~al.}(2010)\citenamefont{Dudek, Edwards,
  Peardon, Richards, and Thomas}}]{Dudek:2010wm}
\bibinfo{author}{\bibfnamefont{J.~J.} \bibnamefont{Dudek}},
  \bibinfo{author}{\bibfnamefont{R.~G.} \bibnamefont{Edwards}},
  \bibinfo{author}{\bibfnamefont{M.~J.} \bibnamefont{Peardon}},
  \bibinfo{author}{\bibfnamefont{D.~G.} \bibnamefont{Richards}},
  \bibnamefont{and} \bibinfo{author}{\bibfnamefont{C.~E.}
  \bibnamefont{Thomas}}, \bibinfo{journal}{Phys.Rev.}
  \textbf{\bibinfo{volume}{D82}}, \bibinfo{pages}{034508}
  (\bibinfo{year}{2010}), \eprint{1004.4930}.

\bibitem[{\citenamefont{Beane et~al.}(2004)\citenamefont{Beane, Bedaque,
  Parreno, and Savage}}]{Beane:2003da}
\bibinfo{author}{\bibfnamefont{S.~R.} \bibnamefont{Beane}},
  \bibinfo{author}{\bibfnamefont{P.~F.} \bibnamefont{Bedaque}},
  \bibinfo{author}{\bibfnamefont{A.}~\bibnamefont{Parreno}}, \bibnamefont{and}
  \bibinfo{author}{\bibfnamefont{M.~J.} \bibnamefont{Savage}},
  \bibinfo{journal}{Phys. Lett.} \textbf{\bibinfo{volume}{B585}},
  \bibinfo{pages}{106} (\bibinfo{year}{2004}), \eprint{hep-lat/0312004}.

\bibitem[{\citenamefont{Lee}(2006)}]{Lee:2005fk}
\bibinfo{author}{\bibfnamefont{D.}~\bibnamefont{Lee}}, \bibinfo{journal}{Phys.
  Rev.} \textbf{\bibinfo{volume}{B73}}, \bibinfo{pages}{115112}
  (\bibinfo{year}{2006}), \eprint{cond-mat/0511332}.

\bibitem[{\citenamefont{Edwards and Joo}(2005)}]{Edwards:2004sx}
\bibinfo{author}{\bibfnamefont{R.~G.} \bibnamefont{Edwards}} \bibnamefont{and}
  \bibinfo{author}{\bibfnamefont{B.}~\bibnamefont{Joo}}
  (\bibinfo{collaboration}{SciDAC Collaboration, LHPC Collaboration, UKQCD
  Collaboration}), \bibinfo{journal}{Nucl.Phys.Proc.Suppl.}
  \textbf{\bibinfo{volume}{140}}, \bibinfo{pages}{832} (\bibinfo{year}{2005}),
  \eprint{hep-lat/0409003}.

\bibitem[{\citenamefont{Clark et~al.}(2010)\citenamefont{Clark, Babich, Barros,
  Brower, and Rebbi}}]{Clark:2009wm}
\bibinfo{author}{\bibfnamefont{M.}~\bibnamefont{Clark}},
  \bibinfo{author}{\bibfnamefont{R.}~\bibnamefont{Babich}},
  \bibinfo{author}{\bibfnamefont{K.}~\bibnamefont{Barros}},
  \bibinfo{author}{\bibfnamefont{R.}~\bibnamefont{Brower}}, \bibnamefont{and}
  \bibinfo{author}{\bibfnamefont{C.}~\bibnamefont{Rebbi}},
  \bibinfo{journal}{Comput.Phys.Commun.} \textbf{\bibinfo{volume}{181}},
  \bibinfo{pages}{1517} (\bibinfo{year}{2010}), \eprint{0911.3191}.

\bibitem[{\citenamefont{Babich et~al.}(2011)\citenamefont{Babich, Clark, Joo,
  Shi, Brower et~al.}}]{Babich:2011np}
\bibinfo{author}{\bibfnamefont{R.}~\bibnamefont{Babich}},
  \bibinfo{author}{\bibfnamefont{M.}~\bibnamefont{Clark}},
  \bibinfo{author}{\bibfnamefont{B.}~\bibnamefont{Joo}},
  \bibinfo{author}{\bibfnamefont{G.}~\bibnamefont{Shi}},
  \bibinfo{author}{\bibfnamefont{R.}~\bibnamefont{Brower}},
  \bibnamefont{et~al.} (\bibinfo{year}{2011}), \eprint{1109.2935}.

\bibitem[{\citenamefont{{The HDF Group}}(1997-NNNN)}]{hdf5}
\bibinfo{author}{\bibnamefont{{The HDF Group}}},
  \emph{\bibinfo{title}{{Hierarchical Data Format, version 5}}}
  (\bibinfo{year}{1997-NNNN}), \bibinfo{note}{http://www.hdfgroup.org/HDF5/}.

\bibitem[{\citenamefont{Kurth et~al.}(2015)\citenamefont{Kurth, Pochinsky,
  Sarje, Syritsyn, and Walker-Loud}}]{Kurth:2015mqa}
\bibinfo{author}{\bibfnamefont{T.}~\bibnamefont{Kurth}},
  \bibinfo{author}{\bibfnamefont{A.}~\bibnamefont{Pochinsky}},
  \bibinfo{author}{\bibfnamefont{A.}~\bibnamefont{Sarje}},
  \bibinfo{author}{\bibfnamefont{S.}~\bibnamefont{Syritsyn}}, \bibnamefont{and}
  \bibinfo{author}{\bibfnamefont{A.}~\bibnamefont{Walker-Loud}},
  \bibinfo{journal}{PoS} \textbf{\bibinfo{volume}{LATTICE2014}},
  \bibinfo{pages}{045} (\bibinfo{year}{2015}), \eprint{1501.06992}.

\end{thebibliography}

\end{document}